\documentclass[aps,prb,twocolumn,showpacs,groupedaddress, amsmath]{revtex4}

\usepackage{graphicx}
\usepackage{amssymb}

\begin{document}


\title{Effect of random disorder and spin frustration on the reentrant spin glass phase and ferromagnetic phase in stage-2 Cu$_{0.93}$Co$_{0.07}$Cl$_{2}$ graphite intercalation compound near the multicritical point}


\author{Itsuko S. Suzuki}
\email[]{itsuko@binghamton.edu}
\affiliation{Department of Physics, State University of New York at 
Binghamton, Binghamton, New York 13902-6000}

\author{Masatsugu Suzuki}
\email[]{suzuki@binghamton.edu}
\affiliation{Department of Physics, State University of New York at 
Binghamton, Binghamton, New York 13902-6000}


\date{\today}

\begin{abstract}
Stage-2 Cu$_{0.93}$Co$_{0.07}$Cl$_{2}$ graphite intercalation compound magnetically behaves like a reentrant ferromagnet near the multicritical point ($c_{MCP} \approx 0.96$). It undergoes two magnetic phase transitions at $T_{RSG}$ ($= 6.64 \pm 0.05$ K) and $T_{c}$ ($= 8.62 \pm 0.05$ K). The static and dynamic nature of the ferromagnetic and reentrant spin glass phase has been studied using DC and AC magnetic susceptibility. Characteristic memory phenomena of the DC susceptibility are observed at $T_{RSG}$ and $T_{c}$. The nonlinear AC susceptibility $\chi_{3}^{\prime}$ has a positive local maximum at $T_{RSG}$, and a negative local minimum at $T_{c}$. The relaxation time $\tau$ between $T_{RSG}$ and $T_{c}$ shows a critical slowing down: $\tau$ with $x = 13.1 \pm 0.4$ and $\tau_{0}^{*} = (2.5 \pm 0.5) \times 10^{-13}$ sec. The influence of the random disorder on the critical behavior above $T_{c}$ is clearly observed: $\alpha = -0.66$, $\beta = 0.63$, and $\gamma = 1.40$. The exponent of $\alpha$ is far from that of 3D Heisenberg model. 
\end{abstract}


\maketitle



\section{\label{intro}INTRODUCTION}
Magnetic phase transitions in reentrant ferromagnets are one of the most intriguing topics which have been extensively studied in recent years.\cite{Gunnarsson1992,Jonason1996a,Jonason1996b,Suzuki2004,Suzuki2005a} The competing ferromagnetic (FM) and antiferromagnetic (AFM) interactions lead to a very peculiar phase diagram characterized by reentrance phenomena. The system undergoes a paramagnetic (PM) phase to FM phase at a ferromagnetic critical temperature $T_{c}$, and a FM phase to a reentrant spin glass (RSG) phase at a reentrant spin glass transition temperature $T_{RSG}$.

Stage-2 Cu$_{c}$Co$_{1-c}$Cl$_{2}$ graphite intercalation compounds (GIC's) is one of reentrant ferromagnets.\cite{Suzuki1998,Suzuki1999a} The structure of these systems is characterized by a staging structure. The Cu$_{c}$Co$_{1-c}$Cl$_{2}$ intercalate layers sandwiched by adjacent graphene sheets are periodically stacked along the $c$ axis. The Cu$_{c}$Co$_{1-c}$Cl$_{2}$ intercalate layer consists of Cu$_{c}$Co$_{1-c}$ layers sandwiched by adjacent Cl layers. Because of the large separation distance between adjacent Cu$_{c}$Co$_{1-c}$ layers, these systems magnetically behave like a quasi 2D random spin system. In each Cu$_{c}$Co$_{1-c}$ layers Cu$^{2+}$ and Co$^{2+}$ ions are randomly distributed on the lattice sites. The spin frustration effect occurs as a result of the competition between the FM interactions (Co$^{2+}$ - Co$^{2+}$ and Co$^{2+}$ - Cu$^{2+}$) and the AFM interaction (Cu$^{2+}$ - Cu$^{2+}$). In a pure statge-2 CuCl$_{2}$ GIC ($c$ = 1) there is another type of spin frustration effect arising from the frustrated nature of the system: the antiferromagnet on the isosceles triangular lattice.\cite{Suzuki1994}

\begin{figure}
\includegraphics[width=7.0cm]{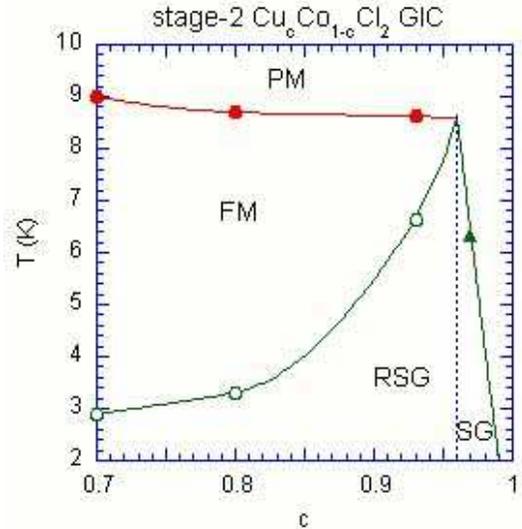}
\caption{\label{fig01}(Color online) Magnetic phase diagram of stage-2 Cu$_{c}$Co$_{1-c}$Cl$_{2}$ GIC for $0.7\le c\le 1$. The multicritical point is located at $c_{MCP} \approx 0.96$ and $T_{MCP} \approx 8.8$ K. The solid lines are guide to the eyes.}
\end{figure}

The magnetic phase diagram of these systems depends on the Cu concentration. The system with $c\ge 0.4$ magnetically behaves like a reentrant ferromagnet (see Fig.~\ref{fig01} for the magnetic phase diagram).\cite{Suzuki1998,Suzuki1999a} The phase transitions occur at a RSG transition temperature ($T_{RSG}$) and a Curie temperature ($T_{c}$). The intermediate phase between $T_{c}$ and $T_{RSG}$ is the FM phase and the low temperature phase is the RSG phase. With further increasing the Cu concentration above $c$ = 0.93, the spin frustration effect is much more enhanced. Above a multicritical point $c_{MCP}$ ($c_{MCP} \approx 0.96$), a FM long range order no longer exists. Only the spin glass phase survives for $c_{MCP}<c<1$. For stage-2 CuCl$_{2}$ GIC ($c = 1$), no phase transition occurs at least above $T$ = 0.3 K, mainly because of the frustrated nature of the 2D antiferromagnet on the isosceles triangular lattice.\cite{Suzuki1994} 

In the present paper, we report our experimental study on the magnetic phase transitions of reentrant ferromagnet, stage-2 Cu$_{0.93}$Co$_{0.07}$Cl$_{2}$ GIC ($T_{RSG}$ = 6.64 K and $T_{c}$ = 8.62 K) near $c=c_{MCP}$ where the PM-FM, FM-RSG and RSG-SG and SG-PM boundaries merge. Because of the enhanced frustrated nature of the system, the FM phase may be very different from an ideal FM phase with long range order. It may be ferromagnetic chaotic phase with short range order.

The static and dynamic nature of the RSG and FM phases is extensively studied from the measurements of DC and AC magnetic susceptibility using a SQUID magnetometer. Our study includes the $T$ dependence of nonlinear AC magnetic susceptibility, the memory phenomena of DC magnetization, the dynamic scaling relation of the absorption $\chi^{\prime\prime}(\omega,T)$, and the scaling plot of $\chi^{\prime}(T,H)$ above $T_{c}$. Detail of the aging dynamics of the present system will be presented elsewhere. For $0.4 \le c \le 0.8$, the absorption of the AC magnetic susceptibility $\chi^{\prime\prime}$ clearly exhibits two peaks at $T_{RSG}$ and $T_{c}$,\cite{Suzuki1999a} while for $c$ = 0.93, no peak in $\chi^{\prime\prime}$ is observed at $T_{RSG}$ (see Sec.~\ref{resultE}). The existence of the RSG phase for $c$ = 0.93 is experimentally confirmed from the above methods.

\section{\label{exp}EXPERIMENTAL PROCEDURE}
The detail of the sample characterization of stage-2 Cu$_{0.93}$Co$_{0.07}$Cl$_{2}$ GIC was reported in our previous paper.\cite{Suzuki1999a} The stoichiometry of the system is given by C$_{n}$Cu$_{0.93}$Co$_{0.07}$Cl$_{2}$ with $n$ = 11.42. Cu$^{2+}$ and Co$^{2+}$ ions are randomly distributed on the triangular lattice sites of the Cu$_{0.93}$Co$_{0.07}$ intercalate layer. The repeat distance between adjacent intercalate layers is $d$ = 12.80 $\pm$ 0.05 {\AA}. The DC magnetization and AC magnetic susceptibility were measured using a SQUID (superconducting quantum interference device) magnetometer (Quantum Design MPMS XL-5) with an ultralow field capability as option. The detail of each experimental procedure will be described in Sec.~\ref{result}. The nonlinear AC magnetic susceptibility was measured as follows, where $h$ is the amplitude of the AC field. After each $h$-scan ($h$ = 1 mOe to 4.2 Oe) at fixed $T$, $T$ was increased by $\Delta T$ = 0.1 K in the temperature range between 1.9 and 12.0 K. The nonlinear AC magnetic susceptibility was measured where $f$ = 1 Hz. The detail of the experimental procedure has been reported in our previous papers.\cite{Suzuki2004,Suzuki2002} Experimentally, $\Theta_{1}^{\prime}$ and $\Theta_{1}^{\prime\prime}$ are the in-phase and out-of phase components of the first harmonics of in the AC magnetization:
\begin{equation} 
\frac{\Theta_{1}^{\prime}}{h}=\chi_{1}^{\prime}+\frac{3}{4}\chi_{3}^{\prime}h^{2}+\frac{10}{16}\chi_{5}^{\prime}h^{4}+\frac{35}{64}\chi_{7}^{\prime}h^{6}+\cdots ,
\label{eq01} 
\end{equation} 
\begin{equation} 
\frac{\Theta_{1}^{\prime\prime}}{h}=\chi_{1}^{\prime\prime}+\frac{3}{4}\chi_{3}^{\prime\prime}h^{2}+\frac{10}{16}\chi_{5}^{\prime\prime}h^{4}+\frac{35}{64}\chi_{7}^{\prime\prime}h^{6}+\cdots .
\label{eq02} 
\end{equation} 
The least squares fits of the data ($\Theta_{1}^{\prime}/h$ vs $h$ and $\Theta_{1}^{\prime\prime}/h$ vs $h$) [10 mOe $\le h \le 4.2$ Oe] for each $T$ yields the nonlinear susceptibility.

\section{\label{result}RESULT}
\subsection{\label{resultA}$T$ dependence of $M_{ZFC}$, $M_{FC}$, $M_{TRM}$, $M_{IRM}$ and $\Delta M$ ($= M_{FC}-M_{ZFC}$)}
The measurements of $M_{ZFC}$, $M_{FC}$, $M_{TRM}$, and $M_{IRM}$ were made as follows. 
(i) \textit{The zero-field cooled magnetization} ($M_{ZFC}$) \textit{measurement}. The system was annealed at 50 K for 1200 sec in the absence of $H$. The system was cooled from 50 to 2 K at $H$ = 0 through the ZFC cooling protocol. After the system was aged at 2 K for $t_{w}$ = 100 sec at $H$ = 0, the magnetic field is applied at $H$ (= 1 Oe) and subsequently $M_{ZFC}$ was measured with increasing $T$ from 2 to 12 K at the rate of 0.025 K/minute. 
(ii) \textit{The field cooled magnetization} ($M_{FC}$) \textit{measurement}. The system was annealed at 50 K for 1200 sec in the presence of $H$. Then the system was quenched from 50 to 12 K in the presence of $H$ through the FC cooling protocol. The magnetization $M_{FC}$ was measured with decreasing $T$ from 12 to 2 K. 
(iii) \textit{The thermoremnant magnetization} ($M_{TRM}$) \textit{measurement}. The system was cooled from 50 to 2 K in the presence of $H$ through the FC cooling protocol. After the system was aged at 2 K for $t_{w}$ = 100 sec, the field was cut off ($H$ = 0). Then the magnetization $M_{TRM}$ was measured with increasing $T$ from 2 to 12 K. 
(iv) \textit{The isothermal remnant magnetization} ($M_{IRM}$) \textit{measurement}. The system was cooled from 50 to 2 K in the absence of $H$ through the ZFC cooling protocol. After the system was aged at 2 K for $t_{w}$ = 100 sec at $H$ = 0. The field was applied at $H$ and then cut off. The magnetization $M_{IR}$ was measured with increasing $T$ from 2 to 12 K at $H$ = 0.

\begin{figure}
\includegraphics[width=7.0cm]{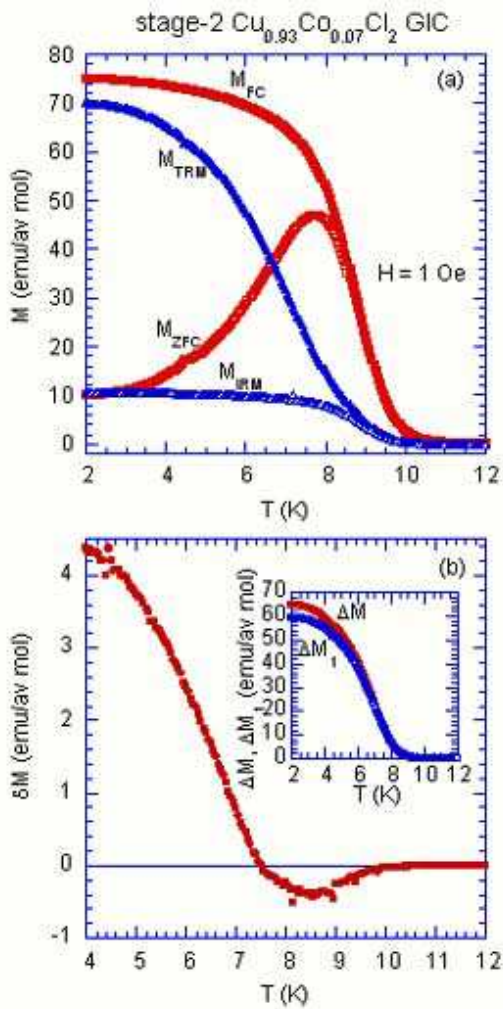}
\caption{\label{fig02}(Color online) (a) $T$ dependence of $M_{FC}$, $M_{ZFC}$, $M_{TRM}$, and $M_{IRM}$ at $H$ = 1 Oe. (b) $T$ dependence of the difference $\delta M=\Delta M-\Delta M_{1}$, where $\Delta M=M_{FC}-M_{ZFC}$. $\Delta M_{1}=M_{TRM}-M_{IRM}$. The $T$ dependence of $\Delta M$ and $\Delta M_{1}$ is shown in the inset.}
\end{figure}

\begin{figure*}
\includegraphics[width=14.0cm]{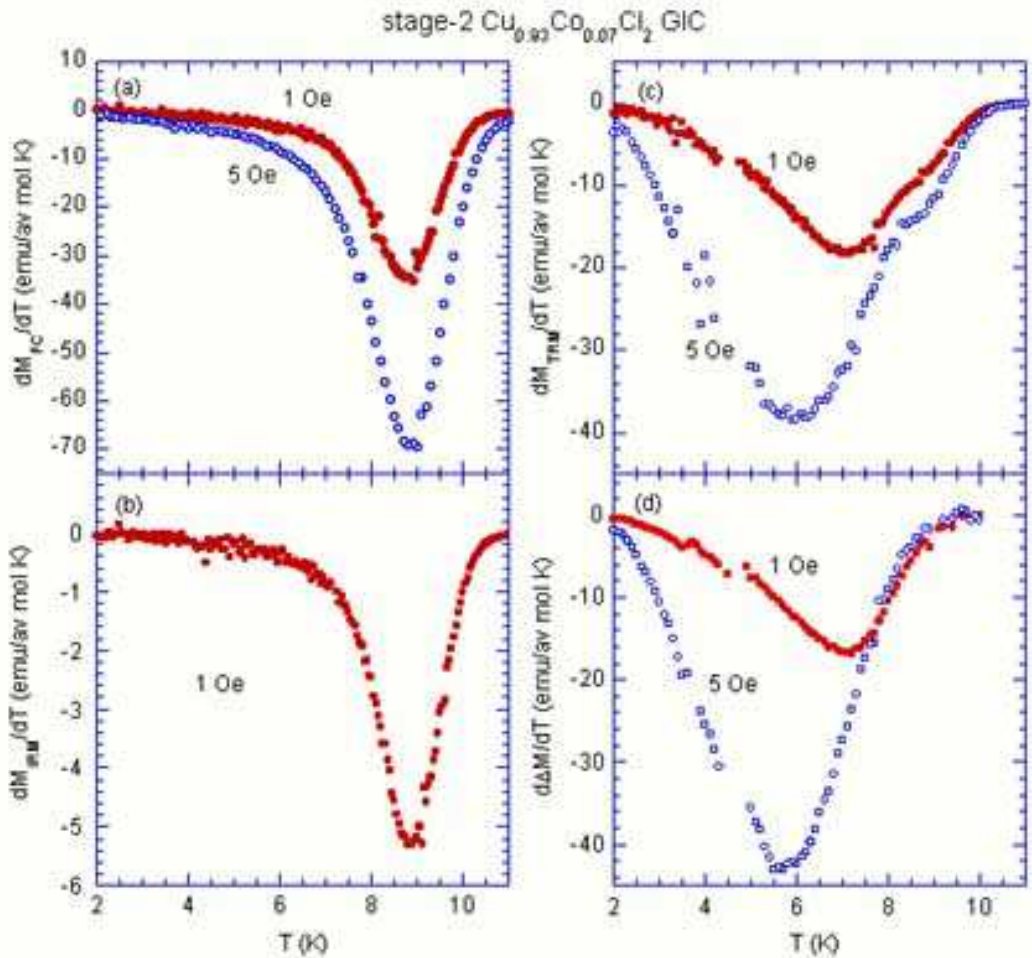}
\caption{\label{fig03}(Color online) $T$ dependence of d$M_{FC}$/d$T$, d$M_{IRM}$/d$T$, d$M_{TRM}$/d$T$, and d$\Delta M$/d$T$. $H$ = 1 and 5 Oe. $\Delta M=M_{FC}-M_{ZFC}$.}
\end{figure*}

\begin{figure}
\includegraphics[width=7.0cm]{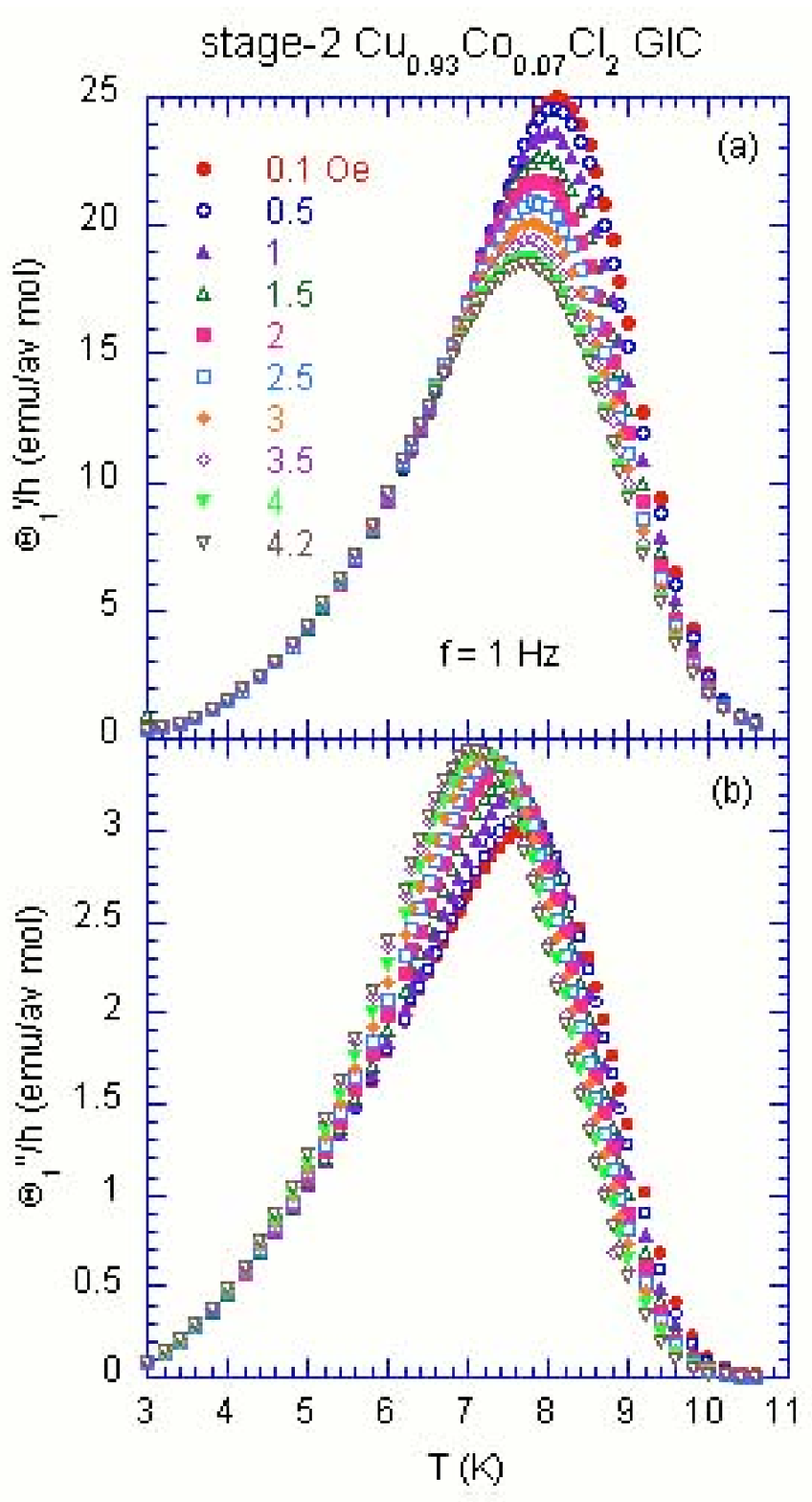}
\caption{\label{fig04}(Color online) $T$ dependence of (a) the dispersion $\Theta _{1}^{\prime}/h$ and (b) the absorption $\Theta_{1}^{\prime\prime}/h$ at various $h$ (= 0.1 - 4.2 Oe). $H$ = 0. $f$ = 1 Hz.}
\end{figure}

Figure \ref{fig02} shows the $T$ dependence of $M_{ZFC}$, $M_{FC}$, $M_{TRM}$, $M_{IRM}$, and $\Delta M=M_{FC}-M_{ZFC}$ at $H$ = 1 Oe. $M_{ZFC}$ shows a peak at 7.75 K. The deviation of $M_{ZFC}$ from $M_{FC}$ starts to occur below about 9.8 K, which is rather higher than the peak temperature of $\chi_{ZFC}$. Figure \ref{fig03} shows the $T$ dependence of the derivatives d$M_{FC}$/d$T$,  d$M_{IRM}$/d$T$, d$M_{TRM}$/d$T$ and d$\Delta M$/d$T$, where $H$ = 1 Oe. The derivatives d$M_{FC}$/d$T$, and d$M_{IRM}$/d$T$ show a negative local minimum near $T_{c}$ (= 8.80 K), while the derivatives d$M_{TRM}$/d$T$ and d$\Delta M$/d$T$ show a negative local minimum near $T_{RSG}$ (= 6.64 k). These results suggest that the system undergoes two magnetic phase transitions at $T_{RSG}$ and $T_{c}$. The inset of Fig.~\ref{fig02} shows the $T$ dependence of $\delta M=\Delta M-\Delta M_{1}$, where $\Delta M_{1}=M_{TRM}-M_{IRM}$. The difference $\delta M$ is positive below 7.5 K. It is negative between 7.5 and 11 K, showing a negative local minimum at 8.5 K near $T_{c}$.\cite{Chantrell1991}

Similar experiments have been carried out at $H$ = 5 Oe. The results are as follows. Both the derivative d$M_{TRM}$/d$T$ and d$\Delta M$/d$T$ exhibit a negative local minimum at $T_{RSG}(H=5$ Oe$) \approx 6$ K, which is indicative of the decrease of $T_{RSG}(H)$ with increasing $H$ from $H$ = 1 to 5 Oe. In contrast, the derivative d$M_{FC}$/d$T$ exhibits a local maximum at $T_{c}(H=5$ Oe$) = 8.8$ K, indicating a slight increase of $T_{c}(H)$ with increasing $H$ from 1 to 5 Oe.

The aging dynamics of this system will be reported elsewhere.\cite{Suzuki2005b} Here we note that the relaxation rate $S(t)$ (= d$\chi_{ZFC}(t)$/d$\ln t$) exhibits a peak at a peak time $t_{cr}$. We find that the peak height of $S(t)$ at $t=t_{cr}$ shows a local maximum at $T_{RSG}$, but no anomaly at $T_{c}$.

\subsection{\label{resultB}Nonlinear AC magnetic susceptibility}

\begin{figure}
\includegraphics[width=7.0cm]{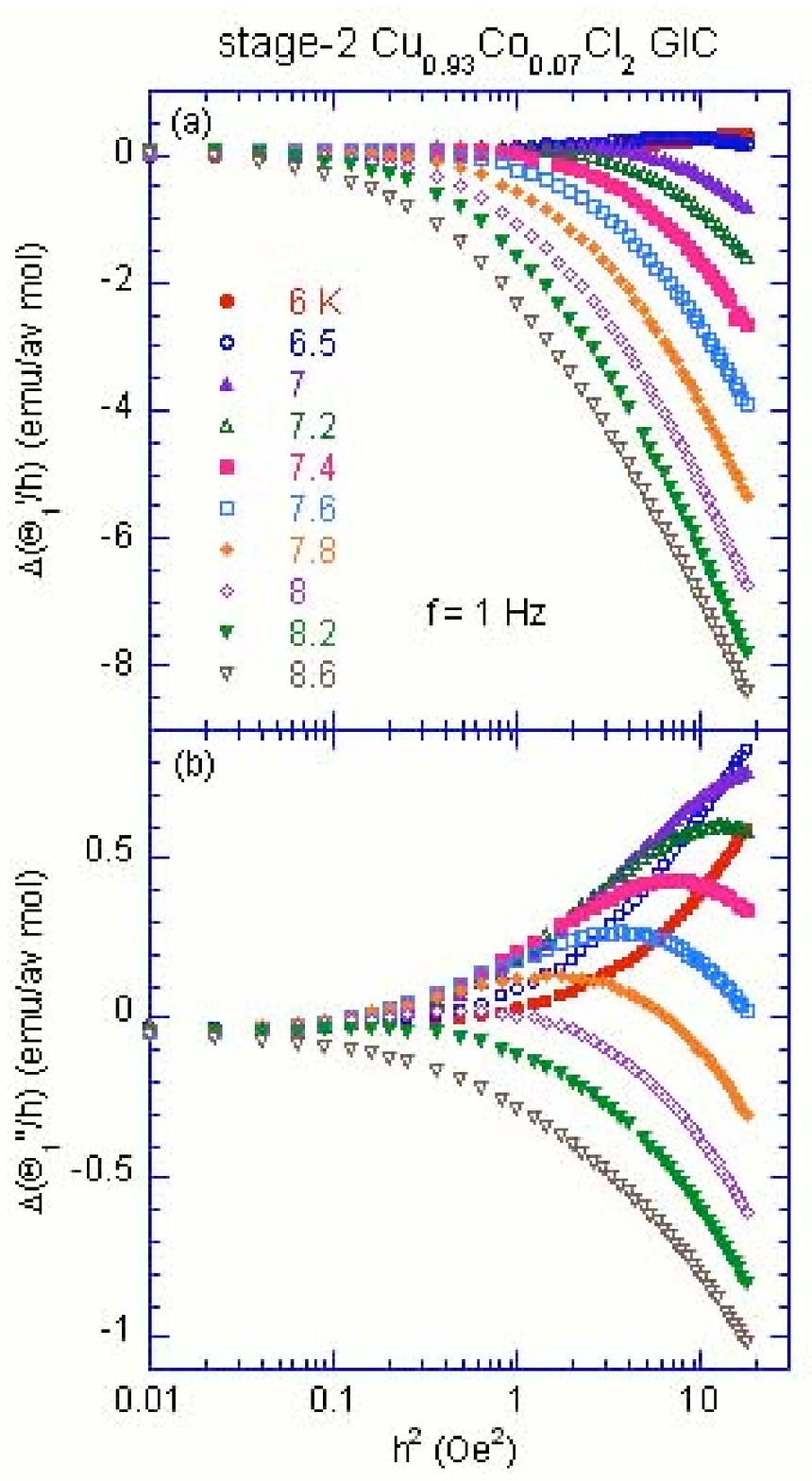}
\caption{\label{fig05}(Color online) Plot of $\Delta (\Theta _{1}^{\prime}/h)$ and $\Delta (\Theta_{1}^{\prime\prime}/h)$ as a function of $h^{2}$, where $\Delta (\Theta_{1}^{\prime}/h)$ and $\Delta (\Theta_{1}^{\prime\prime}/h)$ are defined as the difference between $\Theta_{1}^{\prime}/h$ and $\Theta_{1}^{\prime\prime}/h$ at $h$ and those at $h$ = 0.01 Oe, respectively.}
\end{figure}

Figure \ref{fig04} shows the $T$ dependence of the dispersion ($\Theta_{1}^{\prime}/h$) and the absorption ($\Theta_{1}^{\prime\prime}/h$) of the AC magnetic susceptibility, where $f$ = 1 Hz. The different curves correspond to different amplitude of the AC field, 1 mOe $\le h\le 4.2$ Oe. The features of $\Theta_{1}^{\prime}/h$ vs $T$ and $\Theta_{1}^{\prime\prime}/h$ vs $T$ are summarized as follows. Both $\Theta_{1}^{\prime}/h$ and $\Theta_{1}^{\prime\prime}/h$ exhibit a peak at a temperature between $T_{RSG}$ and $T_{c}$. These peaks linearly shift to the low-$T$ side with increasing the AC amplitude $h$. The curve of $\Theta_{1}^{\prime}/h$ vs $T$ is independent of $h$ below 6 K, but it is strongly dependent on $h$ at temperatures between 6 and 10 K even far above $T_{c}$. The curve of $\Theta_{1}^{\prime\prime}/h$ vs $T$ is independent of $h$ below 5 K but it is strongly dependent on $h$ at temperatures between 5 and 10 K. 

\begin{figure}
\includegraphics[width=7.0cm]{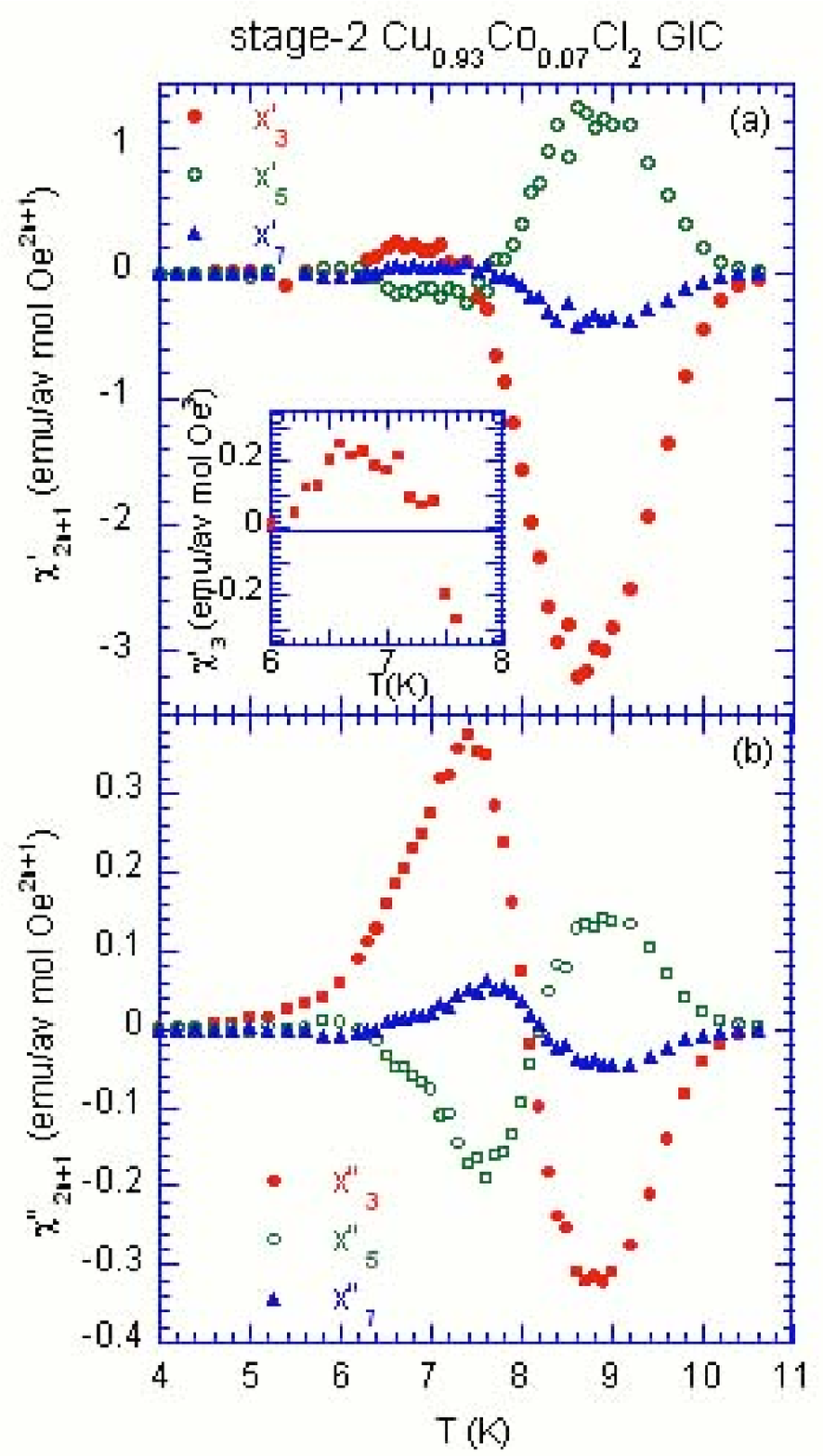}
\caption{\label{fig06}(Color online) $T$ dependence of the nonlinear AC susceptibility. $f$ = 1 Hz. (a) $\chi_{2n+1}^{\prime}$ ($n$ = 1, 2, and 3) (the detail of $\chi_{3}^{\prime}$ vs $T$ is shown in the inset) and (b) $\chi_{2n+1}^{\prime\prime}$ ($n$ = 1, 2, and 3).}
\end{figure}

Both the FM phase and the RSG phase exhibit strong nonlinearities and slow dynamics. There is a characteristic AC field $h_{0}(T)$. For $h<h_{0}(T)$, both $\Theta _{1}^{\prime}/h$ vs $T$ and $\Theta_{1}^{\prime\prime}/h$ vs $T$ do not depend on $h$ below $T_{RSG}$, suggesting the linear response of the system. Here we discuss the nonlinear AC magnetic susceptibility. For convenience, we define $\Delta (\Theta_{1}^{\prime}/h)$ and $\Delta (\Theta_{1}^{\prime\prime}/h)$ as the difference between $\Theta_{1}^{\prime}/h$ and $\Theta_{1}^{\prime\prime}/h$ at $h$ and those at $h$ = 0.01 Oe. Figures \ref{fig05}(a) and (b) show the plot of $\Delta (\Theta_{1}^{\prime}/h)$ and $\Delta (\Theta_{1}^{\prime\prime}/h)$ as a function of $h^{2}$, respectively. Both $\Delta (\Theta_{1}^{\prime}/h)$ and $\Delta (\Theta_{1}^{\prime\prime}/h)$ are strongly dependent on $h^{2}$ between $T_{c}$ and $T_{RSG}$. Figures \ref{fig06}(a) and (b) show the $T$ dependence of the nonlinear AC magnetic susceptibility ($\chi_{3}^{\prime}$, $\chi_{5}^{\prime}$, $\chi_{7}^{\prime}$, $\chi_{3}^{\prime\prime}$, $\chi_{5}^{\prime\prime}$, $\chi_{7}^{\prime\prime}$) at $f$ = 1 Hz. The $T$ dependence of $\chi_{1}^{\prime}$ ($=\chi^{\prime}$) and $\chi_{1}^{\prime\prime}$ ($=\chi^{\prime\prime}$) will be discussed in Sec.~\ref{resultE}. The linear susceptibility ($\chi_{1}^{\prime}$ and $\chi_{1}^{\prime\prime}$) shows no change of sign, while the nonlinear AC susceptibility ($\chi_{3}^{\prime}$, $\chi_{5}^{\prime}$, $\chi_{7}^{\prime}$, $\chi_{3}^{\prime\prime}$, $\chi_{5}^{\prime\prime}$, and $\chi_{7}^{\prime\prime}$) undergoes several changes of sign between 6 and 10 K. The features of the $T$ dependence of the linear and nonlinear AC susceptibilities are summarized as follows. The linear dispersion $\chi_{1}^{\prime}$ exhibits a peak at 8.10 K. The nonlinear dispersion  $\chi_{3}^{\prime}$ starts to appear above $T$ = 6.0 K as $T$ increases and exhibits a positive local maximum at $T_{RSG}$ and a local negative minimum at $T_{c}$, which is much more pronounced compared with the positive local maximum. The nonlinear dispersion $\chi_{5}^{\prime}$ shows a positive local maximum at 6.0 K, a zero crossing at 6.28 K ($<T_{RSG}$), a negative local minimum at 7.42 K, and a positive local maximum at $T_{c}$. The linear absorption $\chi_{1}^{\prime\prime}$ has a peak at 7.75 K. The nonlinear absorption  $\chi_{3}^{\prime\prime}$ starts to appear above $T$ = 5 K with increasing $T$, and exhibits a positive local maximum at 7.42 K between $T_{RSG}$ and $T_{c}$ and a negative local minimum around $T_{c}$.

It is predicted from the mean field theory that the nonlinear DC susceptibility $\chi_{3}$ diverges on both sides of $T_{c}$ for the PM-FM transition of the FM system.\cite{Chikazawa1981} The sign of $\chi_{3}$ changes from negative to positive sign as $T$ decreases and crosses $T_{c}$. On the other hand, $\chi_{3}$ diverges negatively at $T_{SG}$ for the PM-SG transition of the SG system.\cite{Suzuki1977} As far as we know, there has been no theoretical prediction for $\chi_{3}$ vs $T$ for the FM-RSG transition. In our system, the sign of the local maximum in $\chi_{3}^{\prime}$ at $T_{RSG}$ is opposite to that of the local minimum in $\chi_{3}$ at $T_{SG}$ in the SG system. On the other hand the sign of the local minimum in $\chi_{3}^{\prime}$ at $T_{c}$ is the same as that of the local minimum in $\chi_{3}$ at $T_{c}$ in the FM system. Here we note that similar phenomena are also observed in $\chi_{3}^{\prime}$ and $\chi_{3}^{\prime\prime}$ for the reentrant ferromagnet Ni$_{77}$Fe$_{1}$Mn$_{22}$.\cite{Sato22001} The nonlinear dispersion $\chi_{3}^{\prime}$ exhibits a negative local minimum at $T_{RSG}$ and a positive local maximum at $T_{c}$. The nonlinear absorption $\chi_{3}^{\prime\prime}$ exhibits only a negative local minimum at $T_{c}$. No anomaly in $\chi_{3}^{\prime\prime}$ is observed. The sign of $\chi_{3}^{\prime}$ for Ni$_{77}$Fe$_{1}$Mn$_{22}$ is opposite to that of our system. 

In summary, $\chi_{3}^{\prime}$ and $\chi_{3}^{\prime\prime}$ of reentrant ferromagnets show complicated $T$ dependence in the vicinity of $T_{RSG}$ and $T_{c}$. The sign and the position of local minimum and local maximum are not sufficiently understood in terms of the simple mean field theory. This suggests that the $T$ dependence of $\chi_{3}^{\prime}$ and $\chi_{3}^{\prime\prime}$ provides a strong measure for the degree of frustrated nature of the systems near the multicritical point. Further discussion will be presented in Sec.~\ref{disD}.

\subsection{\label{resultC}Memory phenomena in $M_{ZFC}$ and $M_{TRM}$}
We have measured two-types of peculiar memory phenomena for the ZFC and TRM magnetization, which have been found in the stage-2 CoCl$_{2}$ GIC by Matsuura et al.\cite{Matsuura1987} Here we present our result on memory phenomena of $M_{ZFC}$ and $M_{TRM}$ for stage-2 Cu$_{0.93}$Co$_{0.07}$Cl$_{2}$ GIC, which is observed in a series of heating and cooling processes. Such a characteristic phenomenon has been predicted theoretically in spin glass based on a successive bifurcation model of the energy level scheme below the spin freezing temperature.\cite{Matsuura1987b} 

\begin{figure}
\includegraphics[width=7.0cm]{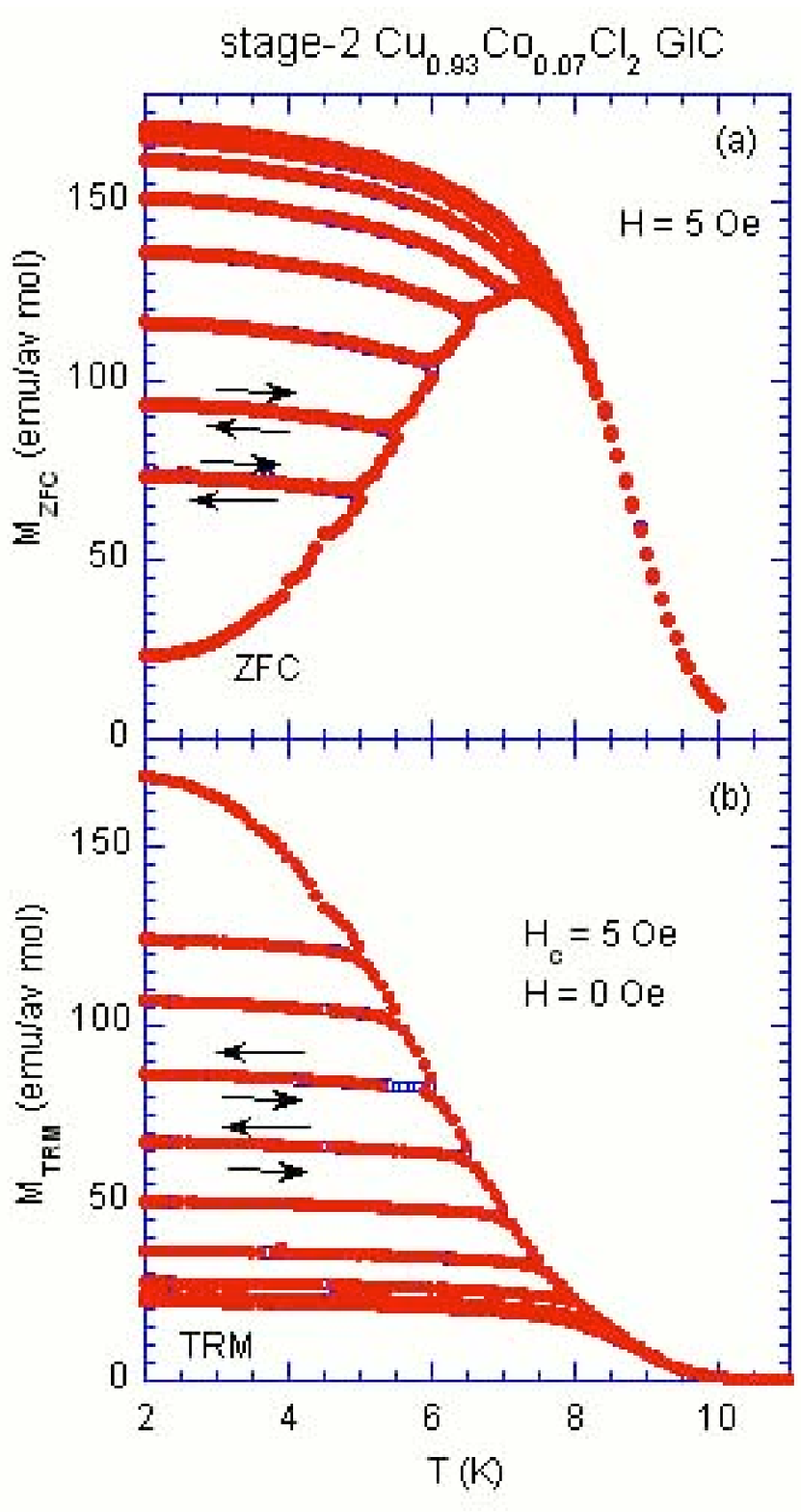}
\caption{\label{fig07}(Color online) (a) $T$ dependence of $M_{ZFC}$ in a series of heating and cooling processes described in the text, after the ZFC cooling protocol from 50 to 2 K. $H$ = 5 Oe. (b) $T$ dependence of $M_{TRM}$ in a series of heating and cooling processes described in the text, after the FC cooling protocol from 50 to 2 K in the presence of $H_{c}$ = 5 Oe. $H$ = 0 during the measurement of $M_{TRM}$.}
\end{figure}

\begin{figure}
\includegraphics[width=6.5cm]{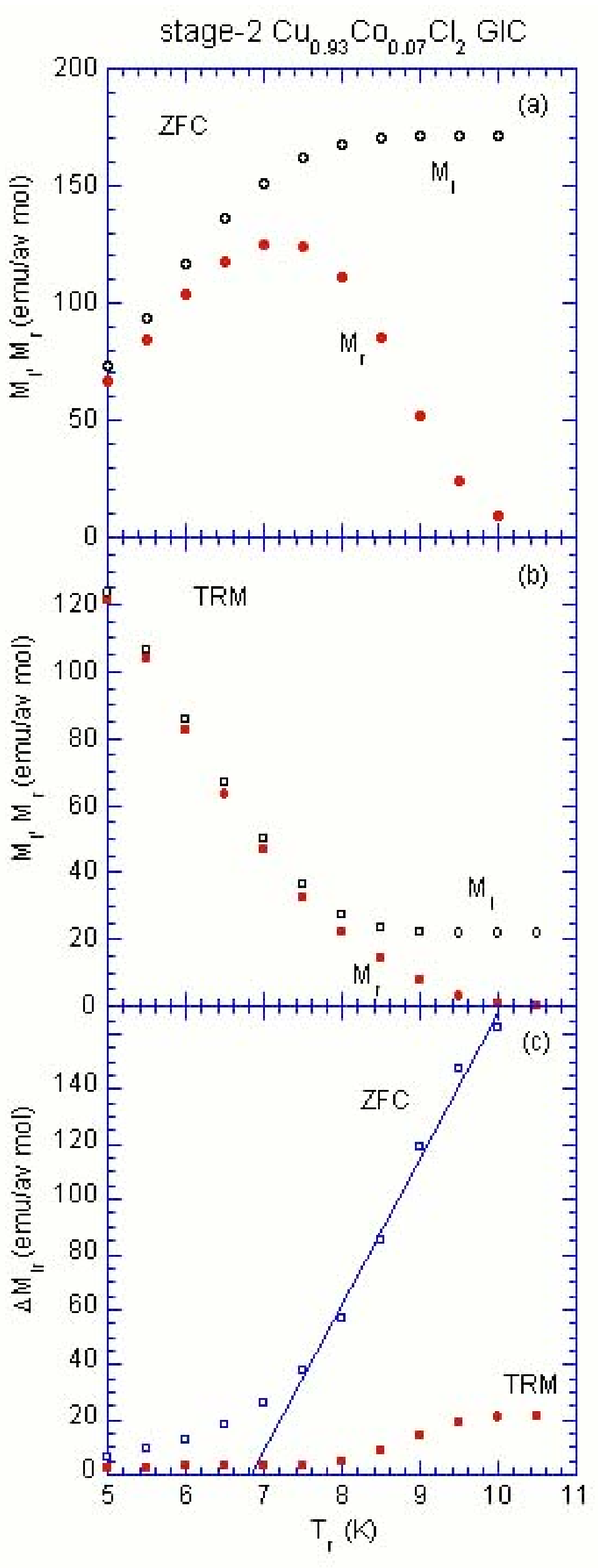}
\caption{\label{fig08}(Color online) (a) $T_{r}$ dependence of $M_{r}$ and $M_{i}$, derived from the measurement of $M_{ZFC}$ with decreasing $T$ from $T_{r}$ (an U-turn temperature) to $T_{i}$ (= 2 K): $M_{i}=M_{ZFC}(T_{i})$ and $M_{r}=M_{ZFC}(T_{r})$. (b) $T_{r}$ dependence of $M_{r}$ and $M_{i}$, derived from the measurement of $M_{TRM}$ with decreasing $T$ from $T_{r}$ to $T_{i}$. $M_{i}=M_{TRM}(T_{i})$ and $M_{r}=M_{TRM}(T_{r})$. (c) $T_{r}$ dependence of $\Delta M_{ir}=M_{i}-M_{r}$ for both the ZFC and TRM cases.}
\end{figure}

(i) \textit{ZFC case}. Before the ZFC magnetization measurement, a zero-field cooling (ZFC) protocol was carried out. It consists of the following processes, (a) annealing of the system at 50 K for 1200 sec in the absence of $H$, (b) quenching of the system from 50 K to 2 K, and (c) aging the system at $T_{i}$ = 2 K and $H$ = 0 for a wait time $t_{w}$ = 100 sec. Just after the magnetic field ($H$ = 5 Oe) is applied to the system, the ZFC magnetization $M_{ZFC}$ was measured with increasing $T$ from $T_{i}$ (= 2 K) to $T_{1}$ (= 5 K) and subsequently with decreasing $T$ from $T_{1}$ to $T_{i}$. Next it was measured with increasing $T$ from $T_{i}$ to $T_{2}$ (= 5.5 K) (the heating process) and subsequently with decreasing $T$ from $T_{2}$ to $T_{i}$ (the cooling process). This process was repeated for the U-turn temperatures $T_{r}$ ($r$ = 3 - 11), where $T_{r}>T_{i}$, $\Delta T=T_{r+1}-T_{r}$ = 0.5 K and $T_{11}$ = 10.5 K. Figure \ref{fig07}(a) shows a typical example of the $T$ dependence of $M_{ZFC}$ using the above method. Note that the value of $M_{ZFC}$ lies between those of $M_{ZFC}^{ref}$ and $M_{FC}^{ref}$ at any $T$ below $T_{c}$. Here $M_{ZFC}^{ref}$ is measured with increasing $T$ from $T_{i}$ to 12 K at $H$ = 5 Oe after the ZFC cooling protocol. The magnetization $M_{FC}^{ref}$ is measured with decreasing $T$ from a temperature far above $T_{c}$ to $T_{i}$ in the presence of $H$ (= 5 Oe). For $T_{r}<T_{RSG}$, the value of $M_{ZFC}$ at $T_{i}$ obtained after the cooling process ($T=T_{r} \rightarrow T_{i}$) is slightly larger than that at $T_{r}$ before the cooling process. The path of $M_{ZFC}$ vs $T$ in the cooling process ($T = T_{r} \rightarrow T_{i}$) is exactly the same as that in the subsequent heating process ($T$ = $T_{i} \rightarrow T_{r}$), indicating the reversibility of such a series of process. The spin configuration imprinted at $T_{r}$ remains unchanged after the cooling and heating processes ($T=T_{r} \rightarrow T_{i} \rightarrow T_{r}$), indicating a memory phenomenon. Even for $T_{RSG}<T<T_{c}$, the path of $M_{ZFC}$ vs $T$ in the cooling process ($T=T_{r} \rightarrow T_{i}$) still coincides with that in the heating process ($T=T_{i} \rightarrow T_{r}$). For $T_{i}\ge T_{c}$, both the path of $M_{ZFC}$ in the cooling process ($T=T_{r} \rightarrow T_{i}$) and path of $M_{ZFC}$ in the heating process ($T=T_{i} \rightarrow T_{r}$) coincide with that of $M_{FC}^{ref}$ which is obtained by cooling from the PM phase to $T=T_{i}$ in the presence of $H$ = 5 Oe. Figure \ref{fig08}(a) shows the $T_{r}$ dependence of $M_{i}$ [$= M_{ZFC}(T_{i})$] and $M_{r}$ [$= M_{ZFC}(T_{r})$] derived from the measurement of $M_{ZFC}$ with decreasing $T$ from $T_{r}$ to $T_{i}$. In Fig.~\ref{fig08}(c) we show the $T_{r}$ dependence of the difference $\Delta M_{ir}^{ZFC}$ ($=M_{i}-M_{r}$). The difference $\Delta M_{ir}^{ZFC}$ starts to increase with increasing $T_{r}$ above $T_{RSG}$. The curve of $\Delta M_{ir}^{ZFC}$ vs $T_{r}$ for $T_{r}>T_{RSG}$ is well described by a straight line, which crosses the $T_{r}$ axis line with $\Delta M_{ir}^{ZFC}$ = 0 around $T_{r}=T_{RSG}$.

(ii) \textit{TRM case}. Before the TRM magnetization measurement, a field cooling (FC) protocol was carried out, consisting of (a) annealing of the system at 50 K for 1200 sec in the presence of $H$ (= 5 Oe), (b) quenching of the system from 50 to 2 K, and (c) aging the system at $T$ = $T_{i}$ = 2 K and $H$ = 5 Oe for a wait time $t_{w}$ = 100 sec. Just after the magnetic field was turned off, the TRM magnetization was measured using the same procedure of heating and cooling: $T_{i} \rightarrow T_{1} \rightarrow T_{i} \rightarrow T_{2} \rightarrow T_{i} \rightarrow T_{3} \rightarrow T_{i} \rightarrow$ and so on. Figure \ref{fig07}(b) shows a typical example of the $T$ dependence of $M_{TRM}$ using this method. The value of $M_{TRM}$ lies between those of $M_{TRM}^{ref}$ and $M_{IRM}^{ref}$ at any $T$ below $T_{c}$. Here the magnetization $M_{TRM}^{ref}$ is measured with increasing $T$ from T$_{i}$ to 12 K at $H$ = 0 after the FC cooling protocol at a field $H_{c}$ = 5 Oe. The magnetization $M_{IRM}^{ref}$ is measured with increasing $T$ from T$_{i}$ to 12 K at $H$ = 0 after the ZFC cooling protocol from 12 to $T_{i}$, switching $H$ from 0 to 5 Oe, aging the system at $H$ = 5 Oe for a wait time $t_{w}$ = 100 sec, and again switching $H$ from 5 Oe to 0. For $T_{r}<T_{RSG}$, the value of $M_{TRM}$ at $T_{i}$ obtained after the cooling process ($T$ = $T_{r} \rightarrow T_{i}$) is nearly equal to that at $T_{r}$. The path of $M_{TRM}$ vs $T$ in the cooling process ($T$ = $T_{r} \rightarrow T_{i}$) coincides with that in the heating process ($T$ = $T_{i} \rightarrow T_{r}$), indicating that the spin configuration at $T_{r}$ is maintained during the cooling and heating process between $T_{r}$ and $T_{i}$. Even for $T_{RSG}<T<T_{c}$, the path of $M_{TRM}$ vs $T$ in the cooling process ($T$ = $T_{r} \rightarrow T_{i}$) is the same as that in the heating process ($T$ = $T_{i} \rightarrow T_{r}$). For $T_{r}\ge T_{c}$, the path of $M_{TRM}$ vs $T$ in the cooling process ($T=T_{r} \rightarrow T_{i}$) is the same as that of $M_{TRM}$ vs $T$ in the heating process ($T=T_{i} \rightarrow T_{r}$). In fact, $M_{TRM}$ in the cooling process ($T=T_{r} \rightarrow T_{i}$) corresponds to $M_{FC}$ which is obtained from the FC cooling under a small remnant field ($\approx 5$ mOe) from the PM phase. Figure \ref{fig08}(b) shows the $T_{r}$ dependence of $M_{i}$ [$=M_{TRM}(T_{i})$] and $M_{r}$ [$=M_{TRM}(T_{r})$] derived from the measurement of $M_{TRM}$ with decreasing $T$ from $T_{r}$ to $T_{i}$. In Fig.~\ref{fig08}(c) we show the $T_{r}$ dependence of the difference $\Delta M_{ir}^{TRM}$ ($=M_{i}-M_{r}$). The difference $\Delta M_{ir}^{TRM}$ is nearly equal to zero for $T_{r}<T_{RSG}$. This suggests that the freezing of the spin configuration occurs due to considerable slowing-down of response to the change of temperature from $T=T_{r}$ to $T_{i}$. It undergoes a step-like change around $T_{r}=T_{RSG}$ and tends to saturate for $T_{r}>T_{c}$. The equilibrated ferromagnetic domains grows during the decrease of $T$ from $T_{r}$ ($T_{RSG}<T_{r}<T_{c}$) to $T_{RSG}$. They becomes frozen in on further cooling below $T_{RSG}$ and is retrieved on reheating up to $T=T_{r}$.

In conclusion, the memory effect observed for $T_{r}<T_{RSG}$ is rather different from that for $T_{RSG}<T_{r}<T_{c}$ for both the ZFC and TRM cases. This clearly demonstrates the difference in the nature of spin order between the RSG phase and the FM phase.

\subsection{\label{resultD}$H$-$T$ phase diagram}
\subsubsection{$\chi^{\prime}$ vs $T$ and $\chi^{\prime\prime}$ vs $T$ in the presence of $H$}

\begin{figure}
\includegraphics[width=6.5cm]{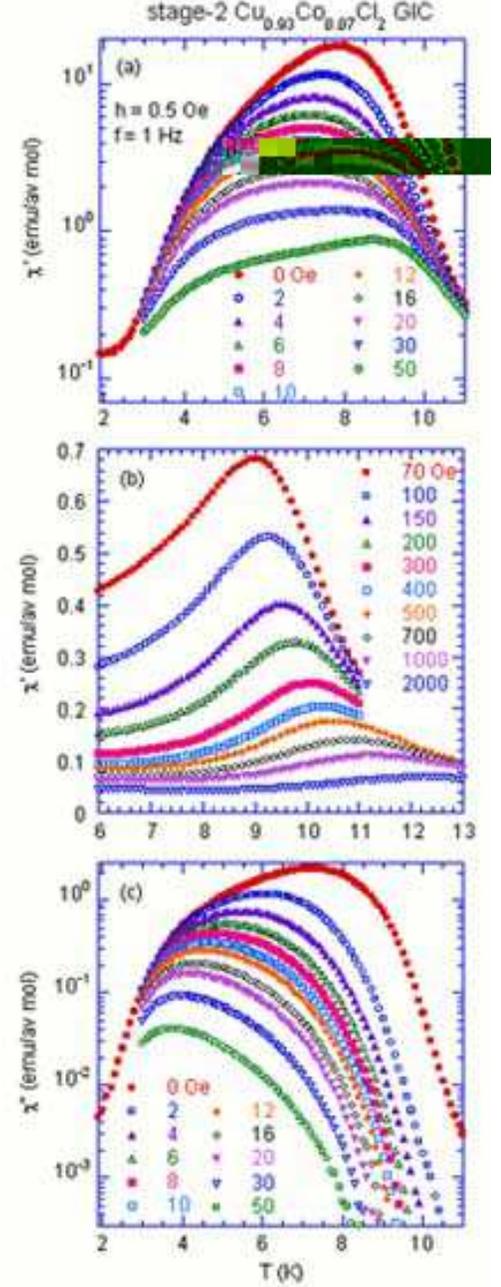}
\caption{\label{fig09}(Color online) $T$ dependence of (a) $\chi^{\prime}$ for $0\le H\le 50$ Oe, (b) $\chi^{\prime}$ for $70\le H\le 2000$ Oe, and (c) $\chi^{\prime\prime}$ for $0\le H\le 50$ Oe. $f$ = 1 Hz. $h$ = 0.5 Oe.}
\end{figure}

\begin{figure}
\includegraphics[width=7.0cm]{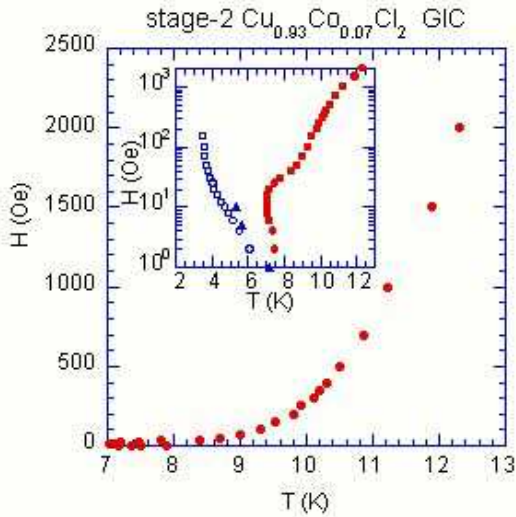}
\caption{\label{fig10}(Color online) $H$-$T$ diagram, where the peak temperatures of $\chi^{\prime}$ vs $T$ ({\Large $\bullet$}) and $\chi^{\prime\prime}$ vs $T$({\Large $\circ$}) are plotted as a function of $T$. Negative local minimum temperatures of d$\Delta\chi$/d$T$ vs $T$ ($\blacktriangle$) are also plotted as a function of $H$. $\Delta\chi = \chi_{FC}-\chi_{ZFC}$.}
\end{figure}

\begin{figure}
\includegraphics[width=7.0cm]{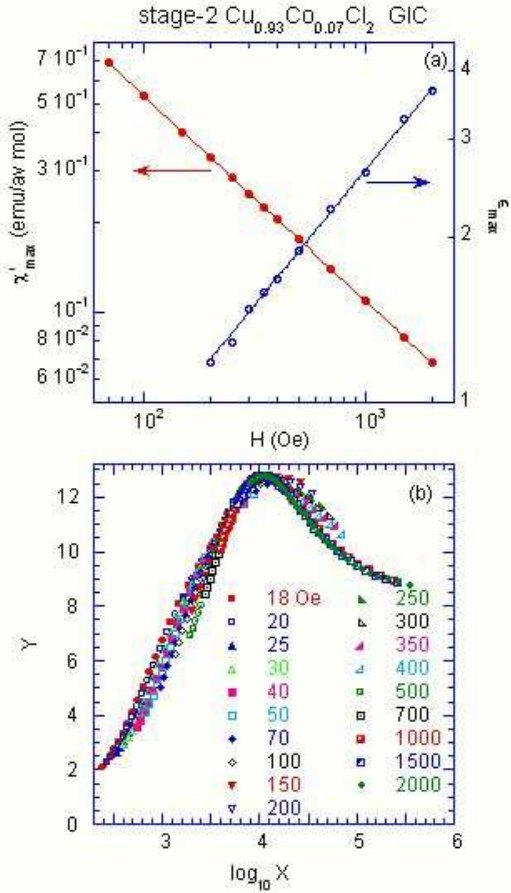}
\caption{\label{fig11}(Color online) (a) $H$ dependence of $\chi^{\prime}_{max}$ and $\epsilon_{max}$ ($= T_{max}/T_{c}-1$), where $T_{c}=8.62$ K. (b) Scaling plot of $Y=\chi^{\prime}(T,H)H^{\gamma /\Delta}$ as a function of $X=H/\epsilon ^{\Delta }$. $\epsilon =T/T_{c}-1$. $T_{c}=8.62$ K. $\Delta$ = 2.03. $9.3\le T\le 13.0$ K.}
\end{figure}

Figures \ref{fig09}(a), (b), and (c) show the $T$ dependence of $\chi^{\prime}$ and $\chi^{\prime\prime}$ in the presence of $H$ ($0 \le H\le 2500$ Oe), where $f$ = 1 Hz and $h$ = 0.5 Oe. The peak of $\chi^{\prime}$ associated with the RSG transition shifts to the low-$T$ side with increasing $H$ for $0\le H\le 10$ Oe. The peak height drastically decreases with increasing $H$. At $H$ = 50 Oe, a broad peak associated with the FM transition becomes pronounced around $T=T_{c}$, because of the strong suppression of the broad RSG peak in the presence of $H$. The FM peak shifts to the high-$T$ side with increasing $H$ above 100 Oe. In contrast, the peak of $\chi^{\prime\prime}$ shifts to the low-$T$ side with increasing $H$ ($0\le H\le 150$ Oe). The peak height drastically decreases with increasing $H$ and reduces to zero above 150 Oe. In Fig.~\ref{fig10}, we show the $H$-$T$ diagram where the peak temperatures of $\chi^{\prime}$ and $\chi^{\prime\prime}$ are plotted as a function of $H$. There are two kinds of transitions at $T_{RSG}(H)$ and $T_{c}(H)$. The RSG phase is suppressed and the FM phase is enhanced by the application of $H$ at least above 20 Oe. The apparent decrease in the peak temperature of $\chi^{\prime\prime}$ vs $T$ with increasing $H$ at low $H$ is due to the decrease in $T_{RSG}(H)$ as a result of the suppression in the RSG contribution of $\chi^{\prime}$. In summary, $T_{RSG}(H)$ decreases with increasing $H$, while $T_{c}(H)$ increases with increasing $H$. Similar behavior in $\chi^{\prime}$ above $T_{c}$ in the presence of $H$ has been observed in a 2D XY-like ferromagnet K$_{2}$CuF$_{4}$.\cite{Suzuki1981} Using the same method used in K$_{2}$CuF$_{4}$, we examine the static scaling hypothesis for the dispersion $\chi^{\prime}(T,H)$ above $T_{c}$ in the presence of $H$. The dispersion $\chi^{\prime}(T,H)$ is described by a scaling function,
\begin{eqnarray}
\chi^{\prime}(T,H)&=&\epsilon^{-\gamma}f(H/\epsilon^{\Delta}) \nonumber \\
&=&(H/\epsilon^{\Delta })^{\gamma/\Delta}H^{-\gamma /\Delta}f(H/\epsilon^{\Delta}) \nonumber \\
&=&H^{-\gamma/\Delta} \psi (H/\epsilon^{\Delta}),
\label{eq03}
\end{eqnarray}
where $\epsilon =T/T_{c}-1$ and $f(H/\epsilon^{\Delta})$ and $\psi (H/\epsilon^{\Delta })=( H/\epsilon^{\Delta })^{\gamma /\Delta}f(H/\epsilon^{\Delta})$ are single-valued functions of $H/\epsilon^{\Delta }$. The critical exponent $\Delta$ is defined as $\Delta =\beta +\gamma$, where $\beta$ and $\gamma$ are critical exponents of magnetization and susceptibility. We define $\chi^{\prime}_{max}$ and $T_{max}$ as the peak height and the peak temperature of the broad peak of $\chi^{\prime}(T,H)$ vs $T$ at the fixed $H$, respectively. $T_{max}$ is considered to be a temperature below which the magnetic-field induced ferromagnetic state appears. Figure \ref{fig11}(a) shows $\chi^{\prime}_{max}$ as a function of $H$. The least-squares fit of the data of $\chi^{\prime}_{max}$ vs $H$ for 70 Oe$\le H\le 2$ kOe to a power law form $\chi^{\prime}_{max } \approx H^{-\gamma /\Delta}$ yields the exponent $\gamma /\Delta = 0.689 \pm 0.001$. In Fig.~\ref{fig11}(a) we also show the deviation $\epsilon_{max}$ ($=T_{max }/T_{c}-1$) as a function of $H$, where $T_{c}$ = 8.62 K. The least-squares fit of the data to a power law form $H=H_{0}\epsilon_{max}^{\Delta }$ for 200 Oe $\le H\le 2$ kOe yields the exponent $\Delta = 2.03 \pm 0.03$ and $H_{0}= 139 \pm 4$ Oe. Using the scaling relation, $\alpha +2\beta +\gamma = 2$, we have $\alpha = -0.66$, $\beta$ = 0.63, and $\gamma$ = 1.40, where $\alpha$ is the critical exponent of the heat capacity. In Fig.~\ref{fig11}(b) we show a scaling plot of $Y=\chi^{\prime}(T,H)H^{\gamma /\Delta}$ as a function of $X=H/\epsilon^{\Delta}$, where all the data of $\chi^{\prime}(T,H)$ vs $T$ with fixed $H$ (70 Oe $\le H\le 2$ kOe and $9.3\le T\le 14$ K) are plotted. Almost all the data points are well located on a scaling function which has a broad peak around $X=10^{4}$, indicating the validity of the static scaling hypothesis for the present system above $T_{c}$ in an external magnetic field. The critical exponent obtained here will be discussed in Sec.~\ref{disC}.

\subsubsection{$\chi_{ZFC}$ vs $T$, $\chi_{FC}$ vs $T$, and $\Delta\chi$ vs $T$ in the presence of $H$}

\begin{figure}
\includegraphics[width=7.0cm]{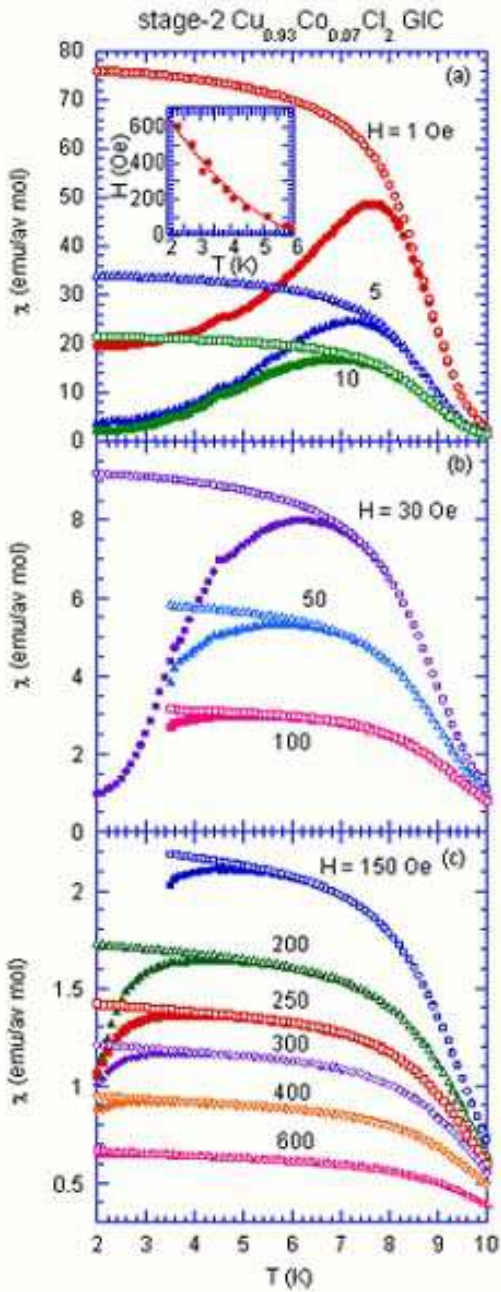}
\caption{\label{fig12}(Color online) (a), (b) and (c) $T$ dependence of $\chi_{FC}$ and $\chi_{ZFC}$. $1\le H\le 600$ Oe. The $H$-$T$ diagram is shown in the inset of (a), where the peak temperatures of $\chi_{ZFC}$ vs $T$ are plotted as a function of $T$.}
\end{figure}

\begin{figure}
\includegraphics[width=7.0cm]{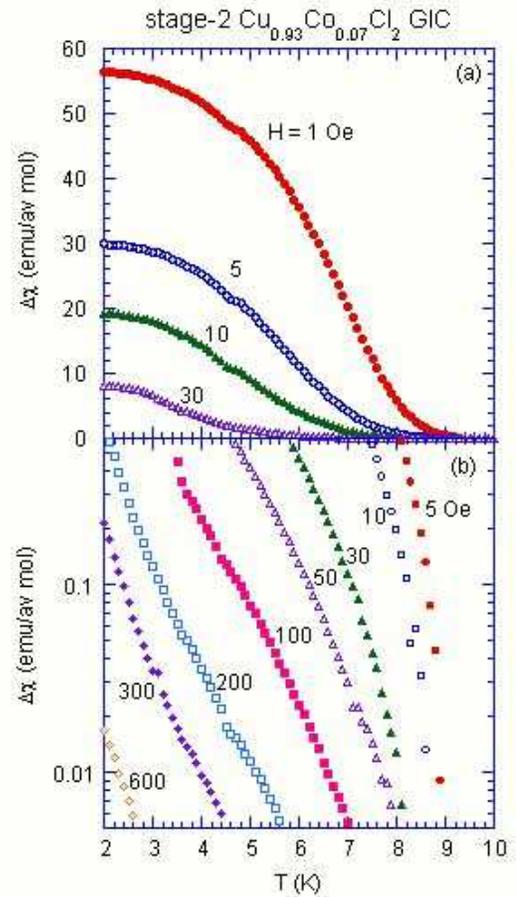}
\caption{\label{fig13}(Color online) (a) and (b) $T$ dependence of $\Delta\chi$ ($=\chi_{FC}-\chi_{ZFC}$) at various $H$.}
\end{figure}

Figure \ref{fig12} shows the $T$ dependence of $\chi_{ZFC}$ and $\chi_{FC}$ in the presence of the fixed $H$ ($H = 1 - 600$ Oe). The susceptibility $\chi_{ZFC}$ shows a broad peak, which shifts to the low-$T$ side with increasing $H$. Correspondingly, the peak height becomes smaller and the peak width becomes broader. The deviation of $\chi_{ZFC}$ from $\chi_{FC}$ occurs below some characteristic temperature dependent on $H$. The susceptibility $\chi_{FC}$ is nearly temperature independent at the lowest $T$. These features are indicative of the existence of the RSG phase below $T_{RSG}(H)$. Here we define $\Delta\chi$ as the difference between $\chi_{FC}$ and $\chi_{ZFC}$: $\Delta\chi =\chi_{FC} -\chi_{ZFC}$. This difference provides a measure for the irreversibility of susceptibility. In Fig.~\ref{fig13} we show the $T$ dependence of $\Delta\chi$ at various $H$. The difference $\Delta\chi$ drastically decreases with increasing $T$. The RSG transition temperature $T_{RSG}(H)$ is usually defined as a temperature at which $\Delta\chi$ is equal to zero. However, it is a little difficult to determine the transition temperature from this definition for the present system. For convenience, we define the RSG transition temperature at which the derivative d$\Delta\chi$/d$T$ has a negative local minimum. The derivative d$\Delta\chi$/d$T$ exhibits a local minimum at $H$ = 1, 5, and 10 Oe. The temperature for the negative local minimum drastically decreases with increasing $T$. The data thus obtained is plotted in the $H$-$T$ diagram (see the inset of Fig.~\ref{fig10}). These data points are located near the line for the $H$ vs $T_{RSG}$ where the peak temperatures of $\chi^{\prime\prime}(T,H)$ ($f$ = 1 Hz and $h$ = 0.5 Oe) is plotted as a function of $H$.

\subsubsection{AT-like transition with an exponent $p$ (= 3/2)}
The inset of Fig.~\ref{fig12}(a) shows the $H$-$T$ diagram where the peak temperature of $\chi_{ZFC}$ is plotted as a function of $H$. This peak temperature at low $H$ is a little higher than $T_{RSG}$ (= 6.64 K). The least-squares fit of the data ($H$ vs $T$) in the limited temperature range ($2.3\le T\le 6$ K) to a power law form
\begin{equation}
H=H_{0}^{*}(1-\frac{T}{T_{RSG}})^{p} ,
\label{eq04}
\end{equation}
yields the exponent $p=1.57\pm 0.12$ and a magnetic field $H_{0}^{*} = 1.16\pm 0.11$ kOe, where $T_{RSG}$ is fixed as $T_{RSG}$ = 6.64 K. The value of the exponent $p$ is close to an AT value predicted by de Almeida and Thouless (AT): $p = 3/2$.\cite{AT1978} This result indicates the SG-like nature of RSG phase for the transition at $T_{RSG}$. Note that in the droplet picture\cite{Fisher1998} the SG transition can be destroyed in the absence of $H$. In fact, this picture is experimentally supported for Fe$_{0.5}$Mn$_{0.5}$TiO$_{3}$\cite{Gunnarsson1991} and Cu$_{0.5}$Co$_{0.5}$Cl$_{2}$-FeCl$_{3}$ graphite bi-intercalation compound (GBIC).\cite{Suzuki2005c}

\subsection{\label{resultE}$f$ and $T$ dependence of $\chi^{\prime}$ and $\chi^{\prime\prime}$}

\begin{figure}
\includegraphics[width=7.0cm]{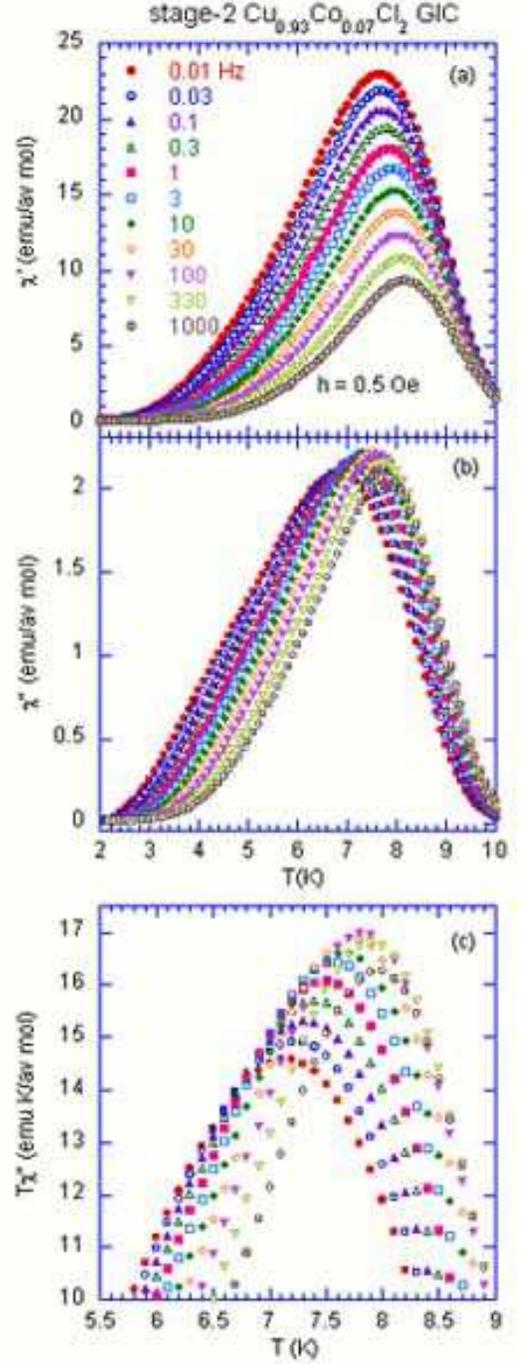}
\caption{\label{fig14}(Color online) $T$ dependence of (a) $\chi^{\prime}$ and (b) $\chi^{\prime\prime}$ at various $f$. $0.01\le f\le 1000$ Hz. $h$ = 0.5 Oe. $H$ = 0 Oe. (c) $T$ dependence of $T\chi^{\prime\prime}(\omega,T)$ at various $f$ ($0.01\le f\le 1000$ Hz). $h$ = 0.5 Oe. $H$ = 0 Oe.}
\end{figure}

Figures \ref{fig14}(a) and (b) show the $T$ dependence of the dispersion $\chi^{\prime}(\omega,T)$ and the absorption $\chi^{\prime\prime}(\omega,T)$ at various $f$ ($0.01\le f\le 1000$ Hz). Since we use the AC field with $h$ = 0.5 Oe, we have $\chi^{\prime} \approx \chi_{1}^{\prime}$ and $\chi^{\prime\prime} \approx \chi_{1}^{\prime\prime}$. The absorption $\chi^{\prime\prime}$ exhibits a peak at $T$ = 6.66 K for $f$ = 0.01 Hz, while the dispersion $\chi^{\prime}$ exhibits a peak at $T$ = 7.65 K for $f$ = 0.01 Hz. The derivative $\partial \chi^{\prime\prime}(\omega ,T)/\partial T$ has a negative local minimum at $T$ = 8.48 K, which corresponds to the temperature of the inflection point. Note that this inflection point does not coincide with the peak temperature of $\chi^{\prime}(\omega,T)$. Both peaks of $\chi^{\prime}(\omega,T)$ and $\chi^{\prime\prime}(\omega,T)$ shift to high $T$-side with increasing $f$. 

\begin{figure}
\includegraphics[width=7.0cm]{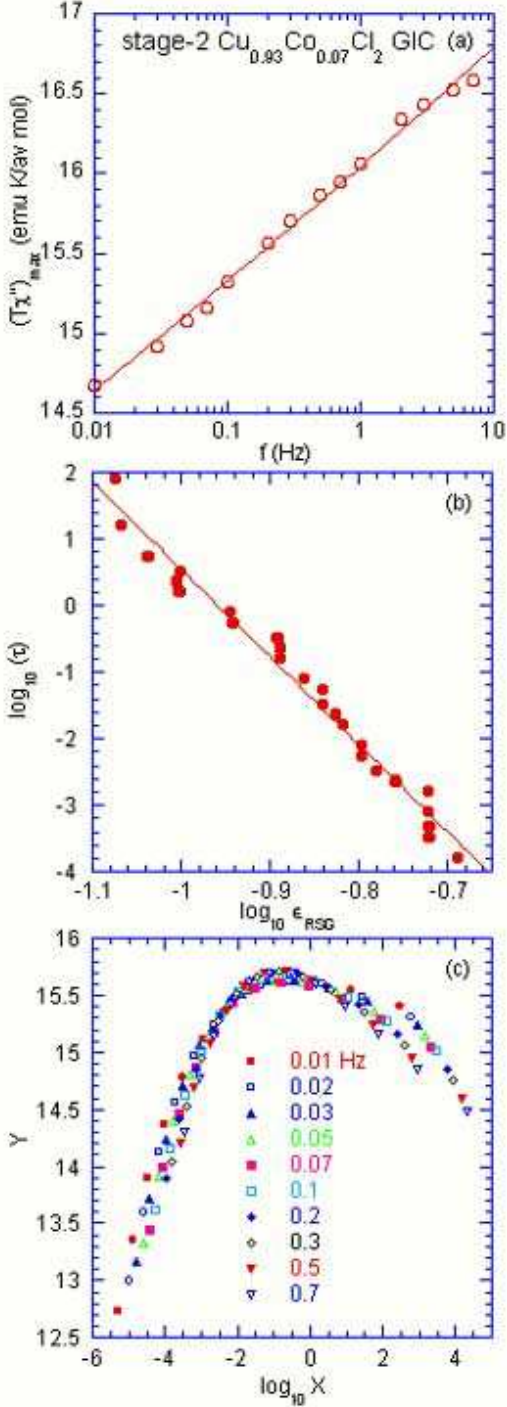}
\caption{\label{fig15}(Color online) (a) Peak height of the data of $T\chi^{\prime\prime}(\omega,T)$ vs $T$, denoted as $[T\chi^{\prime\prime}(\omega,T)]_{max}$ as a function of $f$. $0.01\le f\le 10$ Hz. (b) Plot of log$_{10}\tau$ vs log$_{10}\epsilon_{RSG}$ for $T\chi^{\prime\prime}(\omega,T)$ vs $T$. $\epsilon_{RSG}=T/T_{RSG}-1$. $7.2\le T\le 8.0$ K. $T_{RSG}$ = 6.64 K. The solid lines denote a least-squares fitting curve. The fitting parameters are given in the text. (c) Scaling plot of $Y=T\chi ^{\prime\prime}(\omega,T)/\omega^{\beta _{RSG}/x}$ as a function of $X=\omega\tau$. $0.01\le f\le 0.7$ Hz. $\beta_{RSG}/x$ = 0.0199. The expression of $\tau$ is described in the text.}
\end{figure}

In Fig.~\ref{fig14}(c) we make a plot of $T\chi^{\prime\prime}(\omega,T)$ as a function of $T$. The peak of the curve of $T\chi^{\prime\prime}(\omega,T)$ vs $T$ shifts to the high-$T$ side with increasing $f$. The peak height increases with increasing $f$ in the limited frequency range ($0.01\le f\le 10$ Hz). Here we assume that $T\chi^{\prime\prime}(\omega,T)$ can be described by a dynamic scaling law above $T_{RSG}$;\cite{Rigaux1995}
\begin{eqnarray}
T\chi^{\prime\prime}(\omega ,T)&=&\epsilon^{\beta_{RSG}}G(\omega\tau ) \nonumber \\
&=&(\frac{\omega \tau}{\tau_{0}^{*}})^{-\beta_{RSG}/x}\omega^{\beta_{RSG}/x}G(\omega \tau ) \nonumber \\
&\approx & \omega^{\beta_{RSG}/x}g(\omega\tau ),
\label{eq05}
\end{eqnarray}
where $\omega$ ($=2\pi f$) is the angular frequency, $\beta_{RSG}$ is a critical exponent, and $G(\zeta)$ and $g(\zeta)$ with $\zeta=\omega\tau$ are scaling functions
\begin{equation}
G(\zeta)=\zeta^{\beta_{RSG}/x}g(\zeta),
\label{eq06}
\end{equation}
The relaxation time $\tau$ diverges on approaching $T_{RSG}$ from the high-$T$ side (a conventional critical slowing down),
\begin{equation}
\tau =\tau_{0}^{*}\epsilon^{-x} ,
\label{eq07}
\end{equation}
where $\tau_{0}^{*}$ is a microscopic relaxation time, $\epsilon =T/T_{RSG}-1$, $x=\nu z$, $z$ is a dynamic critical exponent, and $\nu$ is the exponent for the spin correlation length. For simplicity, we assume that the scaling function $g(\omega\tau)$ has a peak at $\omega\tau =1$. In other words, it follows that $T\chi^{\prime\prime}(\omega,T)/\omega^{\beta_{RSG}/x}$ exhibits a peak at the peak temperature at which $\omega\tau =1$. Figure \ref{fig15}(a) shows the peak height of the data of $T\chi^{\prime\prime}(\omega,T)$ vs $T$, denoted as $[T\chi^{\prime\prime}(\omega,T)]_{max}$, as a function of $f$. The peak height increases with increasing $f$ for $0.01\le f\le 100$ Hz, showing a local maximum at $f$ = 100 Hz, and decreases with further increasing $f$. The least squares fit of the data ($[T\chi^{\prime\prime}(\omega,T)]_{max}$ vs $f$) to a power law form ($\omega^{\beta_{RSG}/x}$) for $0.01\le f\le 10$ Hz yields the exponent $\beta_{RSG}/x = 0.0199\pm 0.0004$. This value of $\beta_{RSG}/x$ is much smaller than that ($= 0.071 \pm 0.005$) reported for the reentrant Ising spin glass Fe$_{0.62}$Mn$_{0.38}$TiO$_{3}$.\cite{Mattsson1995} Figure \ref{fig15}(b) shows the $T$ dependence of the relaxation time $\tau$, where $\tau = 1/\omega$ and $T$ is the peak temperature of $T\chi^{\prime\prime}(\omega,T)$ vs $T$. The relaxation time $\tau$ drastically increases with decreasing $T$. The least-squares fit of the data of $\tau$ vs $T$ for $0.01\le f\le 10$ Hz and $6.8\le T\le 8$ K to Eq.(\ref{eq07}) yields the parameters $T_{RSG} = 6.64 \pm 0.1$ K, $x= 13.1 \pm 0.4$ and $\tau_{0}^{*}=(2.5\pm 0.5) \times 10^{-13}$ sec. In Fig.~\ref{fig15}(b) we show the plot of log$_{10}(\tau)$ vs log$_{10}(T/T_{RSG}-1)$ with $T_{RSG}$ = 6.64 K, where the solid line denotes the best-fit line. The value of $x$ in the present system is very close to that of a reentrant Ising spin glass Fe$_{0.62}$Mn$_{0.38}$TiO$_{3}$ ($x = 13\pm 2$).\cite{Mattsson1995} Thus the transition from the FM phase to the RSG phase is dynamically similar to an ordinary transition from the PM phase to the SG phase in spin glass systems. Figure \ref{fig15}(c) shows a scaling plot of $Y=T\chi^{\prime\prime}(\omega,T)/\omega^{\beta_{RSG}/x}$ as a function of $X=\omega\tau$, where $0.01\le f\le 10$ Hz , $\beta_{RSG}/x= 0.0199$, and $T_{RSG}$ = 6.64 K. It seems that almost all the data fall well on an unique scaling function of $\omega\tau$, which has a very broad peak centered at $\omega\tau$ = 0.1, partly because of the broad distribution of the relaxation times over the system. The critical exponent $\beta_{RSG}$ is estimated as $\beta_{RSG} = 0.25 \pm 0.02$. This value of $\beta_{RSG}$ is smaller than those of $\beta_{RSG}$ for the reentrant ferromagnet Cu$_{0.2}$Co$_{0.8}$Cl$_{2}$-FeCl$_{3}$ GBIC ($\beta_{RSG}$ = 0.57)\cite{Suzuki2005a} and $\beta_{SG}$ for the 3D Ising spin glass Cu$_{0.5}$Co$_{0.5}$Cl$_{2}$-FeCl$_{3}$ GBIC ($\beta_{SG} = 0.36 \pm 0.04$).\cite{Suzuki2003}

In summary, the feature of the RSG-FM transition for stage-2 Co$_{0.93}$Co$_{0.07}$Cl$_{2}$ GIC is characterized by small $\beta_{RSG}$ (= 0.25) and the divergence of the relaxation time obeying a power law form given by Eq.(\ref{eq07}) with large $x$ (= 13.1), $\tau_{0}^{*} = (2.5 \pm 0.5) \times 10^{-13}$ sec, and $T_{RSG}$ = 6.64 K.

\subsection{\label{resultF}$f$ dependence of $\chi^{\prime}(\omega,T)$ and $\chi^{\prime\prime}(\omega,T)$ vs $T$ at fixed $T$}

\begin{figure}
\includegraphics[width=7.0cm]{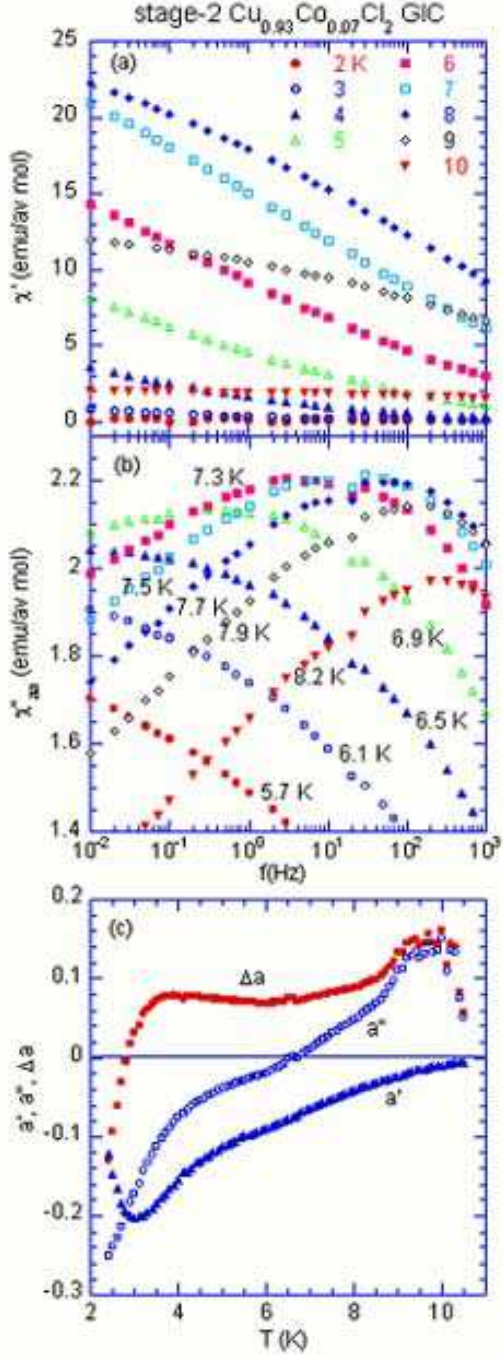}
\caption{\label{fig16}(Color online) $f$ dependence of (a) $\chi^{\prime}$ and (b) $\chi^{\prime\prime}$ at various $T$. $h$ = 0.5 Oe. $H$ = 0 Oe. $T$ = 2.4 - 11 K. (c) $T$ dependence of the exponent $a^{\prime\prime}$, $a^{\prime}$, $\Delta a=a^{\prime\prime}-a^{\prime}$. $\chi^{\prime}(\omega,T) \approx \omega^{a^{\prime}}$ and $\chi^{\prime\prime}(\omega,T) \approx \omega^{a^{\prime\prime}}$ for $0.01\le f\le 0.1$ Hz.}
\end{figure}

Figures \ref{fig16}(a) and (b) show the $f$ dependence of $\chi^{\prime}(\omega, T)$ and $\chi^{\prime\prime}(\omega,T)$ at fixed $T$, where $0.01\le f\le 1000$ Hz and $h$ = 0.5 Oe. The absorption $\chi^{\prime\prime}(\omega,T)$ curves exhibit different characteristics depending on $T$. For $T\le 6.4$ K, $\chi^{\prime\prime}(\omega,T)$ decreases with increasing $f$. For $6.5\le T\le 8.0$ K, $\chi^{\prime\prime}(\omega,T)$ shows a peak at a characteristic frequency, shifting to the low $f$-side as $T$ decreases. For $T\ge 8.1$ K $\chi^{\prime\prime}(\omega,T)$ decreases with increasing $f$. It is predicted from the scaling law given by Eq.(\ref{eq05}) the $\omega$ dependence of $\chi^{\prime\prime}(\omega ,T)/\chi^{\prime\prime}_{\max}$ at the fixed $T$ coincides with that of the scaling function $G(\omega \tau)$ itself, where $\chi^{\prime\prime}_{\max}$ is the maximum of $\chi^{\prime\prime}(\omega,T)$ vs $f$ at the fixed $T$. In contrast, $\chi^{\prime}(\omega,T)$ decreases with increasing $f$ at any $T$.

We find that $\chi^{\prime}(\omega,T)$ and $\chi^{\prime\prime}(\omega,T)$ ($1.9\le T\le 10.6$ K) are well described by power law forms,
\begin{equation}
\chi^{\prime}(\omega,T) \approx \omega^{a^{\prime}} 
\text{ and }
\chi^{\prime\prime}(\omega,T)\approx\omega^{a^{\prime\prime}},
\label{eq09}
\end{equation}
in the limited frequency range ($0.01\le f\le 0.1$ Hz), respectively, where $a^{\prime}$ and $a^{\prime\prime}$ are the temperature-dependent exponents. In Fig.~\ref{fig16}(c) we show the $T$ dependence of $a^{\prime}$, $a^{\prime\prime}$, and $\Delta a=a^{\prime\prime}-a^{\prime}$. We find that the exponent $a^{\prime\prime}$ is equal to zero at $T=T_{RSG}$. Below $T_{RSG}$, $\chi^{\prime\prime}(\omega,T)$ decreases with increasing $f$ for $0.01\le f\le 1000$ Hz, while above $T_{RSG}$ $\chi^{\prime\prime}(\omega,T)$ increases with $f$ at low $f$, exhibiting a peak, and decreases with further increasing $f$. The exponent $a^{\prime\prime}$ shows a broad peak around $T$ = 9.5 K. The derivative d$a^{\prime\prime}$/d$T$ undergoes a drastic change near $T_{c}$ = 8.6 and 4.5 K. According to the fluctuation and dissipation theorem, the magnetic noise power $S(\omega,T)$ is related to $\chi^{\prime\prime}(\omega,T)$ by\cite{Svedlindh1989}
\begin{equation}
S(\omega,T)=4k_{B}T\frac{\chi^{\prime\prime}(\omega,T)}{\omega} 
\approx\omega^{-1+a^{\prime\prime}},
\label{eq10}
\end{equation}
where $k_{B}$ is the Boltzmann constant. Our result of $a^{\prime\prime}$ = 0 only at $T=T_{RSG}$ indicates that the purely $1/f$ character of the noise spectra appears at $T=T_{RSG}$ in the present system.

The exponent $a^{\prime}$ is negative for any $T$ and exhibits a negative local minimum around 3 K. It increases with increasing $T$: $a^{\prime} \approx -0.07$ at $T=T_{RSG}$. It is predicted that $\chi^{\prime\prime}(\omega, T)$ is related to $\chi^{\prime}(\omega,T)$ by a so-called $\pi$/2 rule\cite{Mydosh1993}
\begin{equation}
\chi^{\prime\prime}(\omega,T)
=-\frac{\pi}{2}\frac{\partial\chi^{\prime}(\omega,T)}{\partial\ln\omega} 
=-\frac{\pi}{2}\omega\frac{\partial \chi^{\prime}(\omega,T)}{\partial\omega},
\label{eq11}
\end{equation}
which leads to the relation of $a^{\prime}=a^{\prime\prime}$. In the present system, this relation does not hold valid for low frequencies ($0.01\le f\le 1$ Hz). The difference $\Delta a=a^{\prime\prime}-a^{\prime}$ is equal to $0.07 - 0.08$ for $4\le T\le 7.2$ K. Note that $\Delta a$ = 0 at $T \approx 2.8$ K.

\section{\label{dis}DISCUSSION}
\subsection{\label{disA}Ferromagnetic exchange interaction between Cu$^{2+}$ and Co$^{2+}$}
In the present system Cu$^{2+}$ and Co$^{2+}$ ions are randomly distributed on the triangular lattice. First we show that the intraplanar exchange interaction $J$(Co-Cu) between the nearest neighbor pairs of Cu$^{2+}$ and Co$^{2+}$ ions should be ferromagnetic. To this end, we have measured the $T$ dependence of $\chi_{FC}$ at $H$ = 1 kOe for the present system ($c$ = 0.93). The magnetic susceptibility for $150\le T\le 300$ K obeys a Curie-Weiss law with the effective magnetic moment $P_{eff}(c=0.93)=2.42 \pm 0.03 \mu_{B}$ and the Curie-Weiss temperature $\Theta(c=0.93)=-38.22 \pm 3.10$ K. According to the molecular field theory, the Curie-Weiss temperature $\Theta(c)$ for stage-2 Cu$_{c}$Co$_{1-c}$Cl$_{2}$ GIC can be expressed by\cite{Suzuki2002}
\begin{widetext}
\begin{equation}
\Theta (c)=\frac{c^{2}P_{eff}^{2}(1)\Theta (1)+(1-c)^{2}P_{eff}^{2}(0)\Theta (0)+2\varepsilon c(1-c)\sqrt{|\Theta (1)\Theta (0)|} P_{eff}(1)P_{eff}(0)}{cP_{eff}^{2}(1)+(1-c)P_{eff}^{2}(0)} ,
\label{eq12}
\end{equation}
\end{widetext}
respectively, where $\Theta(0) = 23.2$ K and $P_{eff}(0) = 5.54 \mu_{B}$ for stage-2 CoCl$_{2}$ GIC and $\Theta(1) = -100.9$ K, $P_{eff}(1) = 2.26 \mu_{B}$ for stage-2 CuCl$_{2}$ GIC.\cite{Suzuki1994} The exchange interaction $J$(Cu-Co) may be expressed by a form
\begin{equation}
J(\text{Cu-Co})=\varepsilon \sqrt{|J(\text{Cu-Cu})J(\text{Co-Co})|},
\label{eq13}
\end{equation}
where $J$(Co-Co) [$= \Theta(0)/3 = 7.73$ K] and $J$(Cu-Cu) [$= \Theta(1)/3 = -33.63$ K] are the intraplanar exchange interactions between the nearest neighbor (Cu$^{2+}$-Cu$^{2+}$) ion pairs and (Co$^{2+}$-Co$^{2+}$) ion pairs, respectively, and $\varepsilon$ is only a parameter to be determined. From Eq.(\ref{eq12}) with $c$ = 0.93 and $\Theta(c=0.93) = -38.22\pm 3.10$ K, the parameter $\varepsilon$ can be uniquely determined as $\varepsilon = 2.26\pm 0.02$, which leads to $J$(Cu-Co) = 36.50 K. Note that the magnitude of FM interaction $J$(Cu-Co) is almost the same as that of AFM interaction $J$(Cu-Cu). This result suggests that the competition between mainly these two interactions gives rise to the spin frustration effect at $c$ = 0.93, forming a model equivalent to $\pm J$ spin glass model. The sign of $\Theta(c)$ changes at $c_{0} = 0.848$ from positive to negative with increasing Cu concentration. This Cu concentration $c_{0}$ is a little smaller than $c_{MCP}$ ($\approx 0.96$). 

\subsection{\label{disB}Magnetic phase diagram ($c$ vs $T$)}
It has been theoretically predicted that the spin frustration plays an important role in 2D antiferromagnets on the triangular lattice (AFT). The phase transition of the AFT model depends on the nature of the spin symmetry. When interactions are restricted to nearest-neighbor spins, the AFT Ising model shows no phase transition at any temperature because of a degeneracy of the ground state caused by spin frustration.\cite{Wannier1950} For Heisenberg symmetry,\cite{Kawamura1984} the AFT model predicts a more complex phase transition, driven by the dissociation of pairs of vortices formed of chirality vectors.

The stage-2 CuCl$_{2}$ GIC ($c$ = 1) magnetically behaves like a quasi 2D Heisenberg-like antiferromagnet ($S$ = 1/2) on the isosceles triangular lattice. Because of fully frustrated nature of the system, no magnetic phase transition is observed at least above 0.3 K.\cite{Suzuki1994} In the weak dilution limit ($c \approx 1$), there occurs another type of spin frustration effect due to the competition between the FM interaction $J$(Cu-Co) and the AFM interaction $J$(Cu-Cu), in addition to the spin frustration effect of the 2D AFT type. 

These spin frustration effects lead to a complicated magnetic phase diagram ($c$ vs $T$). For $c$ = 0.97 a spin glass phase appears below $T_{SG}$ = 6.35 K. For $c$ = 0.93, the system undergoes two magnetic phase transitions at $T_{RSG}$ (= 6.64 K) and $T_{c}$ (= 8.62 K). The magnetic phase diagram for $c\ge 0.4$ consists of the PM, RSG, FM and SG phases. The PM-FM line, the FM-RSG line, the SG-PM line intersect a multicritical point (MCP) at ($c_{MCP}$, $T_{MCP}$), where $c_{MCP} \approx 0.96$ and $T_{MCP}$ = 8.8 K. Near the MCP point ($c<c_{MPC}$), there is a subtle interplay between the FM long range order, and the random disorder and spin frustration of the RSG phase. At the present stage, the nature of the SG phase for $c_{MCP}<c<1$ has not been sufficiently examined yet. The shift in the MCP toward the Cu-rich end is the result of the ferromagnetic Cu-Co interaction. If one of Co$^{2+}$ and Cu$^{2+}$ ions is nonmagnetic, the critical temperature reduces to zero at the percolation threshold concentration ($c_{p} = 0.5$) for the 2D triangular lattice. In fact, in stage-2 Co$_{c}$Mg$_{1-c}$Cl$_{2}$ GIC\cite{Suzuki1999b} where Mg is nonmagnetic, the transition temperature associated with the PM-FM transition tends to reduce to zero at $c \approx 0.5$.

\subsection{\label{disC}Anomaly in critical exponent $\alpha$ associated with the PM-FM transition}
We discuss the critical exponents of the reentrant ferromagnets near the MCP along the PM-FM line. The system undergoes two transitions at $T_{c}$ and T$_{RSG}$. As the concentration approaches the MCP, $T_{c}$ becomes closer to $T_{RSG}$ from the higher-$T$ side. It is considered that the critical behavior near the PM-FM transition temperature $T_{c}$ is strongly influenced by the random disorder and spin frustration. In fact, we show that the critical behavior of our system at $T_{c}$ (= 8.62 K) is characterized by the critical exponents $\alpha = -0.66$, $\beta$ = 0.63, and $\gamma$ = 1.40, where $T_{RSG}$ = 6.64 K. These critical exponents are rather different from those predicted form the conventional models (2D Ising and XY models and the 3D Ising and XY models). The exponent $\gamma$ of our system ($\gamma$ = 1.40) is close to that for the 3D Heisenberg ferromagnet ($\alpha = -0.1336$, $\beta$ = 0.3689, $\gamma$ = 1.3960, and $\nu$ = 0.7112)\cite{Campostrini2002}, while the critical exponents $\alpha$ and $\beta$ are rather close to those of the 3D spherical model ($\alpha = -1$, $\beta$ = 0.5, $\gamma$ = 2, and $\delta$ = 5).\cite{Collins1989} 

Similar critical behavior at $T=T_{c}$ has been observed in several reentrant ferromagnets near the MCP. Yeshurun et al.\cite{Yeshurun1981} have studied the critical behavior of amorphous reentrant ferromagnet (Fe$_{1-c}$Mn$_{c}$)$_{75}$P$_{16}$B$_{6}$A$_{l3}$ at $c$ = 0.32 ($c_{MCP}$ = 0.36) along the PM-FM line: $\beta = 0.40\pm 0.03$, $\delta = 5.3\pm 0.3$, where $T_{c}$ = 100 K and $T_{RSG}$ = 38 K. Using the scaling relations $\alpha +2\beta +\gamma =2$ and $\delta =(\beta +\gamma)/\beta$, the exponents $\alpha$ and $\gamma$ are calculated as $\alpha = -0.52$ and $\gamma$ = 1.72. Yeshurun et al.\cite{Yeshurun1981} have also reported the critical exponents of the reentrant ferromagnet (Fe$_{1-c}$Ni$_{c}$)$_{75}$P$_{16}$B$_{6}$A$_{l3}$ at $c$ = 0.80 ($c_{MCP}$ = 0.83) along the PM-FM line at the PM-FM transition; $\alpha = -0.4$, $\beta = 0.40 \pm 0.04$, $\gamma$ = 1.6, and $\delta = 5.0\pm 0.4$, where $T_{c}$ = 90 K and $T_{RSG}$ = 21 K. Pouget et al.\cite{Pouget1994} have examined the critical exponents of the reentrant ferromagnet CdCr$_{(2-2c)}$In$_{(2c)}$S$_{4}$ at $c$ = 0.05 ($c_{MCP}$ = 0.l5), $T_{c}$ = 68.5 K and $T_{RSG}$ = 10.8 K. The system with $c$ = 0 is a 3D Heisenberg ferromagnet. The critical exponents ($\alpha = -0. 01$, $\beta$ = 0.32, $\gamma$ = 1.37, and $\nu$ = 0.70) are compatible with those for the 3D Heisenberg ferromagnet. A drastic change in the critical exponents is observed for the diluted system with $c$ = 0.05 ($\alpha = -0.57$, $\beta$ = 0.30, $\gamma$ = 1.97).

Thus it may be concluded from the above results that a negative large value of the heat capacity exponent $\alpha$ ($= -0.4 - -0.66$) is a feature common to the critical behavior of reentrant ferromagnets near the MPC. In our case, the pure system ($c$ = 1) belongs to the universality class of the 3D Heisenberg model ($\alpha =-0.1336$) in spite of no phase transition. The heat capacity exponent of our system with $c$ = 0.93 is a weakly diluted random system, where only 7 \% of Cu$^{2+}$ ions are randomly replaced by Co$^{2+}$ ions. It seems that our result violates the so-called Harris criterion.\cite{Harris1974} According to this criterion, the dilution should be relevant only if the specific heat exponent $\alpha$ of the pure system is positive. In contrast, when $\alpha <0$, the dilution does not affect the critical behavior. In other words, the critical exponent of the diluted Heisenberg system is the same as that of the pure Heisenberg system. The Harris criterion may not be valid for the systems with competing exchange interactions, where the degree of disorder and spin frustration are greatly enhanced by the dilution. This result is consistent with the flow diagram of the renormalization group theory in the $(c,T)$ plane.\cite{Meo1995} For the dilution of a Heisenberg ferromagnet with no competing interactions, the renormalization group flow at the PM-FM critical line always ends at the critical point of the pure system ($c$ = 1). In the case of competing interactions, the renormalization group flow always ends at a new fixed point at the critical line.

\subsection{\label{disD}Nonlinear susceptibility near the MCP}
We show that the $T$ dependence of the nonlinear AC susceptibility of our system near the MCP system is much more complicated than we expect. The nonlinear dispersion  $\chi_{3}^{\prime}$ for $f$ = 1 Hz is characterized by the zero value below 6 K, a local positive maximum at $T_{RS}$ (= 6.64 K), the change of sign from positive to negative around 7.4 K with increasing $T$, and a local negative minimum at $T_{c}$. The local negative maximum at $T_{c}$ is much more pronounced than the positive local maximum at $T_{RSG}$. 

Here it is interesting to compare our results of $\chi_{3}^{\prime}$ vs $T$ with those of $\chi_{3}$ vs $T$ for the reentrant ferromagnet (Fe$_{1-c}$Mn$_{c}$)$_{75}$P$_{16}$B$_{6}$A$_{l3}$ with $c$ =0.26, 0.30 and 0.32 near the MCP ($c_{MPT}$ = 0.36), which have been reported by Berndt et al.\cite{Berndt1995,Berndt1998} These systems undergo two transitions at $T_{c}$ and $T_{RSG}$. For $c$ = 0.26, the nonlinear DC susceptibility $\chi_{3}$ has a single negative local minimum at $T_{c}$. No anomaly is observed at $T_{RSG}$. For $c$ = 0.30, $\chi_{3}$ consists of a less pronounced negative local minimum at $T_{RSG}$ and a pronounced negative local-minimum at $T_{c}$. In contrast, for $c$ = 0.32, the strength of these anomalies is reversed. The nonlinear susceptibility $\chi_{3}$ consists of a pronounced negative local minimum at $T_{RSG}$ and a less pronounced negative local minimum at $T_{c}$. These results indicate that as the concentration $c$ approaches $c_{MPC}$ from the low-$c$ side, the contribution of the FM-RSG transition at $T_{RSG}$ to $\chi_{3}$ is more significant than that of the PM-FM transition at $T_{c}$ to $\chi_{3}$. Our result of $\chi_{3}^{\prime}$ for $f$ = 1 Hz is qualitatively similar to those for $c$ = 0.30, except for the difference in the sign of the anomalies of $\chi_{3}$ and $\chi_{3}^{\prime}$ around $T_{RSG}$.

\section{CONCLUSION}
We show that the stage-2 Cu$_{0.93}$Co$_{0.07}$Cl$_{2}$ GIC undergoes two magnetic phase transitions at $T_{RSG}$ ($= 6.64\pm 0.05$ K) and $T_{c}$ ($= 8.62 \pm 0.05$ K). The static and dynamic nature of the RSG and FM phases has been extensively studied using various techniques. The nonlinear AC susceptibility $\chi_{3}^{\prime}$ has a positive local maximum at $T_{RSG}$, and a negative local minimum at $T_{c}$. Peculiar memory phenomena for the ZFC and TRM magnetization are observed around $T_{RSG}$ and $T_{c}$. The relaxation time $\tau$ between $T_{RSG}$ and $T_{c}$ shows a critical slowing down: $\tau=\tau_{0}^{*}(T/T_{RSG}-1)^{-x}$ with $x=13.1\pm 0.4$ and $\tau_{0}^{*} = (2.5 \pm 0.5)\times 10^{-13}$ sec. The critical exponent $\beta_{RSG}$ for the RSG phase is $0.25\pm 0.02$, which is much smaller than that predicted for SG phase. The influence of the random disorder on the critical behavior above $T_{c}$ is clearly observed: $\alpha = -0.66$, $\beta$ = 0.63, and $\gamma$ = 1.40. The exponents of $\alpha$ and $\beta$ are close to those of 3D spherical model. The critical temperature $T_{RSG}(H)$ decreases with increasing temperature according to power law form given by Eq.(\ref{eq04}) with the exponent $p=1.57\pm 0.12$. This value of $p$ is close to $p$ = 3/2 predicted by de Almeida and Thouless. 

\begin{acknowledgments}
The authors are grateful to H. Suematsu for providing single crystal of kish graphite and M. Matsuura for invaluable discussion. They would like to thank M. Johnson, B. Olson, T. Shima, C. Vartuli, and T.-Yu Huang for the sample preparation and characterization by using x-ray diffraction.
\end{acknowledgments}


\begin{references}
\bibitem{Gunnarsson1992}K. Gunnarsson, P. Svedlindh, J.-O. Andersson, P. Nordblad, L. Lundgren, H. Aruga Katori, and A. Ito, Phys. Rev. B \textbf{46}, 8227 (1992).
\bibitem{Jonason1996a}K. Jonason, J. Mattsson, and P. Nordblad, Phys. Rev. Lett. \textbf{77}, 2562 (1996).
\bibitem{Jonason1996b}K. Jonason, J. Mattsson, and P. Nordblad, Phys. Rev. B \textbf{53}, 6507 (1996).
\bibitem{Suzuki2004}M. Suzuki and I.S. Suzuki, Phys. Rev. B \textbf{69}, 144424 (2004). See also references therein.
\bibitem{Suzuki2005a}M. Suzuki and I.S. Suzuki, Phys. Rev. B \textbf{71}, 174437 (2005). See also references therein.
\bibitem{Suzuki1998}I.S. Suzuki and M. Suzuki, Solid State Commun. \textbf{106}, 513 (1998).
\bibitem{Suzuki1999a}I.S. Suzuki and M. Suzuki, J. Phys. Condens. Matter \textbf{11}, 521 (1999).
\bibitem{Suzuki1994}M. Suzuki, I.S. Suzuki, C.R. Burr, D.G. Wiesler, N. Rosov, and K. Koga, Phys. Rev. B \textbf{50}, 9188 (1994).
\bibitem{Suzuki2002}M. Suzuki, I.S. Suzuki, and T.-Yu Huang, J. Phys. Condens. Matter \textbf{14}, 5583 (2002).
\bibitem{Chantrell1991}R.W. Chantrell, M. El-Hilo, and K. O'Grady, IEEE Transactions on Magnetism \textbf{27}, 3570 (1991).
\bibitem{Suzuki2005b}M. Suzuki and I.S. Suzuki, unpublished (2005).
\bibitem{Chikazawa1981}S. Chikazawa, C.J. Sandberg, and Y. Miyako, J. Phys. Soc. Jpn. \textbf{50}, 2884 (1981). See also references therein.
\bibitem{Suzuki1977}Masuo Suzuki, Prog. Theor. Phys. \textbf{58}, 1151 (1977).
\bibitem{Sato22001}T. Sato, T. Ando, T. Ogawa, S. Morimoto, and A. Ito, Phys. Rev. B \textbf{64}, 184432 (2001).
\bibitem{Matsuura1987}M. Matsuura, Y. Karaki, T. Yonezawa, and M. Suzuki, Japanese J. Appl. Phys. \textbf{26}, 773 (1987).
\bibitem{Matsuura1987b}M. Matsuura, N. Tanaka, Y. Karaki, and Y. Murakami, Japanese J. Appl. Phys. \textbf{26-S3}, 797 (1987).
\bibitem{Suzuki1981}M. Suzuki and H. Ikeda, J. Phys. Soc. Jpn. \textbf{50}, 1133 (1981).
\bibitem{AT1978}J.R.L. de Almeida and D.J. Thouless, J. Phys. A \textbf{11}, 983 (1978).
\bibitem{Fisher1998}D.S. Fisher and D.A. Huse, Phys. Rev. B \textbf{38}, 373 (1988); \textbf{38}, 386 (1988).
\bibitem{Gunnarsson1991}K. Gunnarsson, P. Svedlindh, P. Nordblad, L. Lundgren, H. Aruga, and A. Ito, Phys. Rev. B \textbf{43}, 8199 (1991).
\bibitem{Suzuki2005c}I.S. Suzuki and M. Suzuki, Phys. Rev. B, in press.
\bibitem{Rigaux1995}C. Rigaux, Ann. Phys. Paris \textbf{20}, 445 (1995).
\bibitem{Mattsson1995}J. Mattsson, T. Jonsson, P. Nordblad, H. Aruga Katori, and A. Ito, Phys. Rev. Lett. \textbf{74}, 4305 (1995).
\bibitem{Suzuki2003}I.S. Suzuki and M. Suzuki, Phys. Rev. B \textbf{68}, 094424 (2003).
\bibitem{Svedlindh1989}P. Svedlindh, K. Gunnarsson, P. Norlblad, L. Lundgren, H. Aruga Katori, and A. Ito, Phys. Rev. B \textbf{4}\textbf{0}, 7162 (1989).
\bibitem{Mydosh1993}J.A. Mydosh, \textit{Spin Glasses-An Experimental Introduction}, Taylor and Francis (London, 1993).
\bibitem{Wannier1950}G.M. Wannier, Phys. Rev. \textbf{79}, 357 (1950).
\bibitem{Kawamura1984}H. Kawamura and S. Miyashita, J. Phys. Soc. Jpn. \textbf{53}, 4138 (1984).
\bibitem{Suzuki1999b}I.S. Suzuki and M. Suzuki, Phys. Rev. B \textbf{59}, 6943 (1999).
\bibitem{Campostrini2002}M. Campostrini, M. Hasenbusch, A. Pelissetto, P. Rossi, and E. Vicari, Phys. Rev. B \textbf{65}, 144520 (2002).
\bibitem{Collins1989}M.F. Collins, \textit{Magnetic Critical Scattering}, Oxford University Press (New York, 1989).
\bibitem{Yeshurun1981}Y. Yeshurun, M.B. Salamon, K.V. Rao, and H.S. Chen, Phys. Rev. B \textbf{24}, 1536 (1981).
\bibitem{Pouget1994}S. Pouget, M. Alba, and M. Nogues, J. Appl. Phys. \textbf{75}, 5826 (1994).
\bibitem{Harris1974}A.B. Harris, J. Phys. C \textbf{7}, 1671 (1974)
\bibitem{Meo1995}M.D.D. Meo, J.D. Reger, and K. Binder, Physica A \textbf{220}, 628 (1995).
\bibitem{Berndt1995}A.G. Berndt, X. Chen, H.P. Kunkel, and G. Williams, Phys. Rev. B \textbf{52}, 10160 (1995).
\bibitem{Berndt1998}A.G. Berndt, H.P. Kunkel, and G. Williams, J. Phys. Condens. Matter \textbf{10}, 8535 (1998).
\end{references}
\end{document}